\documentclass[%
aip,
amsmath,amssymb,
reprint,%
]{revtex4-1}

\usepackage{graphicx}
\usepackage{dcolumn}
\usepackage{bm}
\usepackage{hyperref}
\usepackage[T1]{fontenc}
\usepackage{mathptmx}
\usepackage{xcolor}

\begin{document}


\title{Quantum Entangled Interferometers}

\author{Z. Y. Ou}
 \email{zheyuou@tju.edu.cn; zou@iupui.edu}
\affiliation{College of Precision Instrument and Opto-Electronics Engineering, Key Laboratory of Opto-Electronics Information Technology,
Ministry of Education, Tianjin University, Tianjin 300072, People's Republic of China}
\affiliation{Department of Physics, Indiana University-Purdue University Indianapolis, Indianapolis, IN 46202, USA}
\author{Xiaoying Li}
 \email{xiaoyingli@tju.edu.cn}
\affiliation{College of Precision Instrument and Opto-Electronics Engineering, Key Laboratory of Opto-Electronics Information Technology,
Ministry of Education, Tianjin University, Tianjin 300072, People's Republic of China}

\begin{abstract}
A new type of quantum entangled interferometer was recently realized that employs parametric amplifiers as the wave splitting and recombination elements. The quantum entanglement stems from the parametric amplifiers, which produce quantum correlated fields for probing the phase change signal in the interferometer. This type of quantum entangled interferometer  exhibits some unique properties that are different from traditional beam splitter-based interferometers such as Mach-Zehnder interferometers. Because of these properties, it is superior to the traditional interferometers in many aspects, especially in the phase measurement sensitivity. We will review its unique properties and  applications in quantum metrology and sensing, quantum information,  and quantum state engineering.
\end{abstract}

\maketitle




\section{Introduction}
\label{sect:intro}  

Interferometry, a technique based on wave interference, played a crucial part in the development of fundamental ideas in physics as well as in the technological advances of mankind. It has become an indispensable part in precision measurement and metrology ever since its inception. Most of the physical quantities such as distance, local gravity fields, magnetic fields that can be measured by the interferometric technique are associated with  the phases of the interfering waves. It is the extreme sensitiveness to the phase change in interferometry that leads to the wide applications of the technique in precision measurement and metrology.

\begin{figure}\label{fig:MZ}
\begin{center}
\includegraphics[width=3.2in]{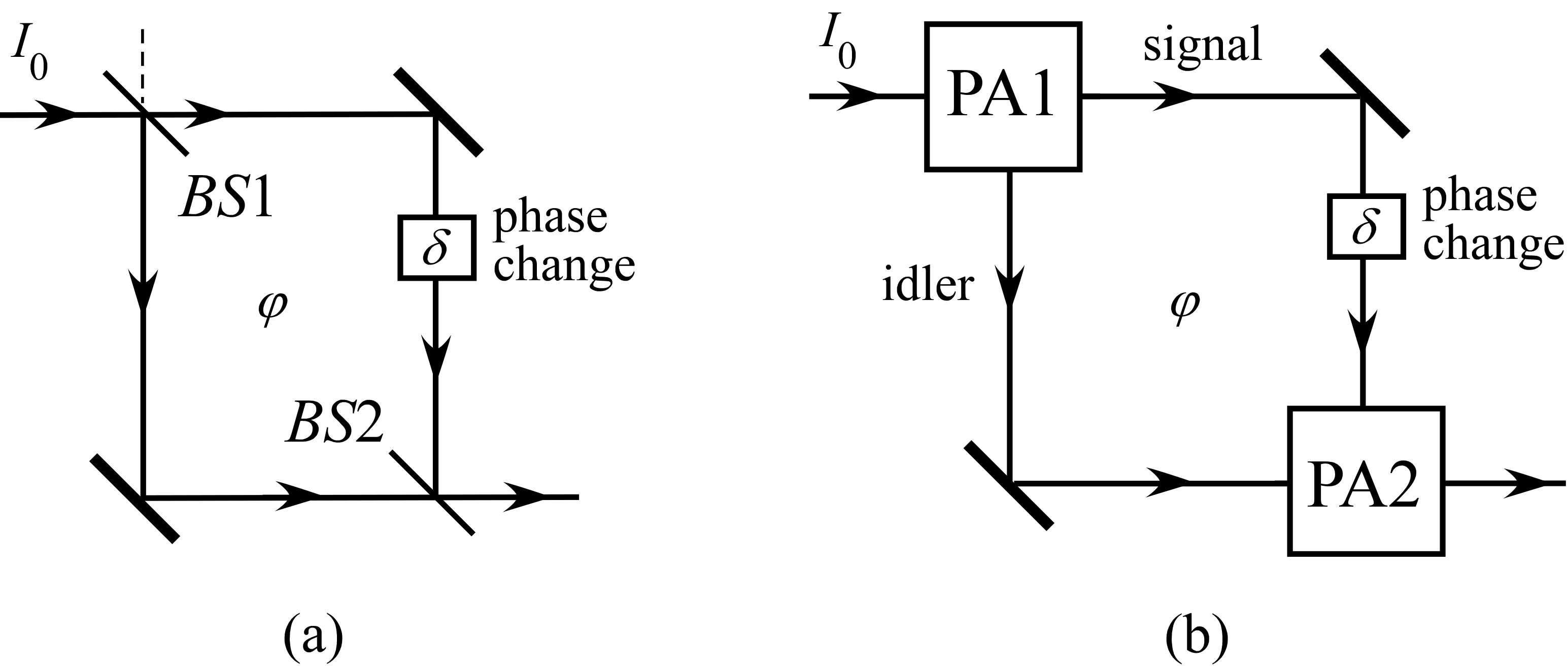}
\end{center}
\caption
{ Comparison between (a) a classical Mach-Zehnder interferometer and (b) an SU(1,1) interferometer. }
\end{figure}

In traditional classical interferometry, as shown in Fig. 1(a), a coherent field is split into two by a beam splitter (BS1). One of the beam, serving as the probe, is phase modulated so as to encode a phase change ($\delta$) on to it. It then interferes with the other beam, serving as a reference, at another beam splitter (BS2), which converts the phase change to an intensity change and the outputs of BS2 are directly measured and analyzed by intensity detectors.
Regardless of difference in design between different schemes, the sensitivity of traditional interferometers is limited by the vacuum quantum noise or the so-called shot noise inherited from injected coherent state and the vacuum state injected from the unused BS input ports \cite{cav} (dashed line in Fig.1(a)) . The sensitivity limit of classical interferometry is referred to as the shot noise limit (SNL) or sometimes the standard quantum limit (SQL).
In order to reduce the vacuum quantum noise, squeezed states are properly injected into interferometers by replacing the vacuum state \cite{cav}.   The result of the squeezed state injection is the reduction of the detection noise below the shot noise level and thus the enhancement of phase measurement sensitivity. Experimental efforts and progresses were made in the generation and application of these quantum states to optical interferometry systems \cite{xiao,gran}.

SU(1,1) interferometers are a new type of quantum interferometers quite different from the classical interferometers in that the linear beam splitters are replaced by nonlinear optical devices of parametric amplifiers, as shown in Fig. 1(b). The name of SU(1,1) stems from the type of interaction involved in parametric processes for coherent wave mixing, which is different from the SU(2)-type interaction for linear wave mixing by a beam splitter. SU(1,1) interferometers were first proposed by Yurke {\it et al.} \cite{yuk} to reach the Heisenberg limit, the ultimate quantum limit in precision phase measurement \cite{ou97}. A modified version with a coherent state  boost \cite{pl,ou12} is more practical to implement experimentally \cite{jing11,hud14}.
Compared to the traditional interferometers with quantum state injections, SU(1,1) interferometers exhibit some distinct features that make them more desirable in practical applications. The first one is that the involvement of nonlinear optical processes for wave mixing allows the coherent superpositions of waves of different types such as atomic spin waves, light waves, and acoustic waves. This type of mixing is impossible for linear beam splitters.
The second is that the employment of parametric amplifiers leads to amplified noise levels at outputs that are much larger than the vacuum noise level. This means the outputs are immune to losses, which are detrimental to quantum information because of the vacuum noise coupled in through the loss channels. The third is that the quantum entanglement generated by parametric amplifiers leads  correlated quantum noise which can be canceled at destructive interference. This gives rise to signal amplification but without noise amplification.

Interference effects involving nonlinear optical processes were demonstrated as soon as the nonlinear optical effects such as second harmonic generation were discovered \cite{Maker}. Because of the involvement of nonlinear optical processes, these nonlinear interference effects have some interesting applications in spectroscopy \cite{ka15}, optical imaging \cite{ku,dev1}, spatial and temporal shaping \cite{isk15-1,isk15-2}. They can be mostly understood with classical wave theory.  At the quantum level of single photons when the gain of the parametric amplifiers is low, interferometers consisting of spontaneous parametric down-conversion were used to study two-photon or multi-photon interference, which cannot be explained by classical theory. These quantum interferometric effects are the basis for optical quantum information sciences \cite{KLM}. All these phenomena were recently reviewed in a comprehensive article \cite{chek}. On the other hand, when the gain of the parametric amplifiers of the SU(1,1) interferometers is high, the quantum noise performance of the interferometers is totally changed. The early research development of the SU(1,1) interferometers in this regime was covered in the comprehensive review article \cite{chek}. But since the publication of the article, there have been many significant progresses in the field, especially in the realization of many variations of the SU(1,1) interferometer and its applications in quantum metrology, quantum information, and quantum state engineering that are not covered by the review article. Furthermore, there are some mis-understanding in the early researches about the working principle of the interferometer for sensitivity improvement, which lead to non-optimized performance. The roles of each nonlinear element in the interferometer were also better understood now, which reveals the underlying physics of the interferometer.

In this paper, we will concentrate on the quantum noise performance of the SU(1,1) interferometers in the high gain regime  with an emphasis on improving the phase measurement sensitivity. We will have an in-depth discussion on the special features of the interferometer in this case, especially on the role played by quantum entanglement. Based on this discussion, we will find the optimum operation conditions for the best performance in the form of phase measurement sensitivity in comparison with the optimized classical interferometers.  We will reveal the difference and similarity between the SU(1,1) interferometers and squeezed state-based traditional interferometers. These are covered in Sects.\ref{sec:II} and \ref{sec:III}. For the experimental implementation of the SU(1,1) interferometer, we will review in Sects.\ref{sec:IV} and \ref{sec:V} the recent realization of different forms of the interferometer including those with different types of waves. We will discuss in \ref{sec:VI} its applications in multi-parameter measurement, quantum information splitting, quantum entanglement measurement, and mode engineering of quantum states. We conclude in \ref{sec:VII} with prospects for future development.

\section{Performance of classical interferometry} \label{sec:II}

\begin{figure}
\begin{center}
\begin{tabular}{c}
\includegraphics[width=3.2 in]{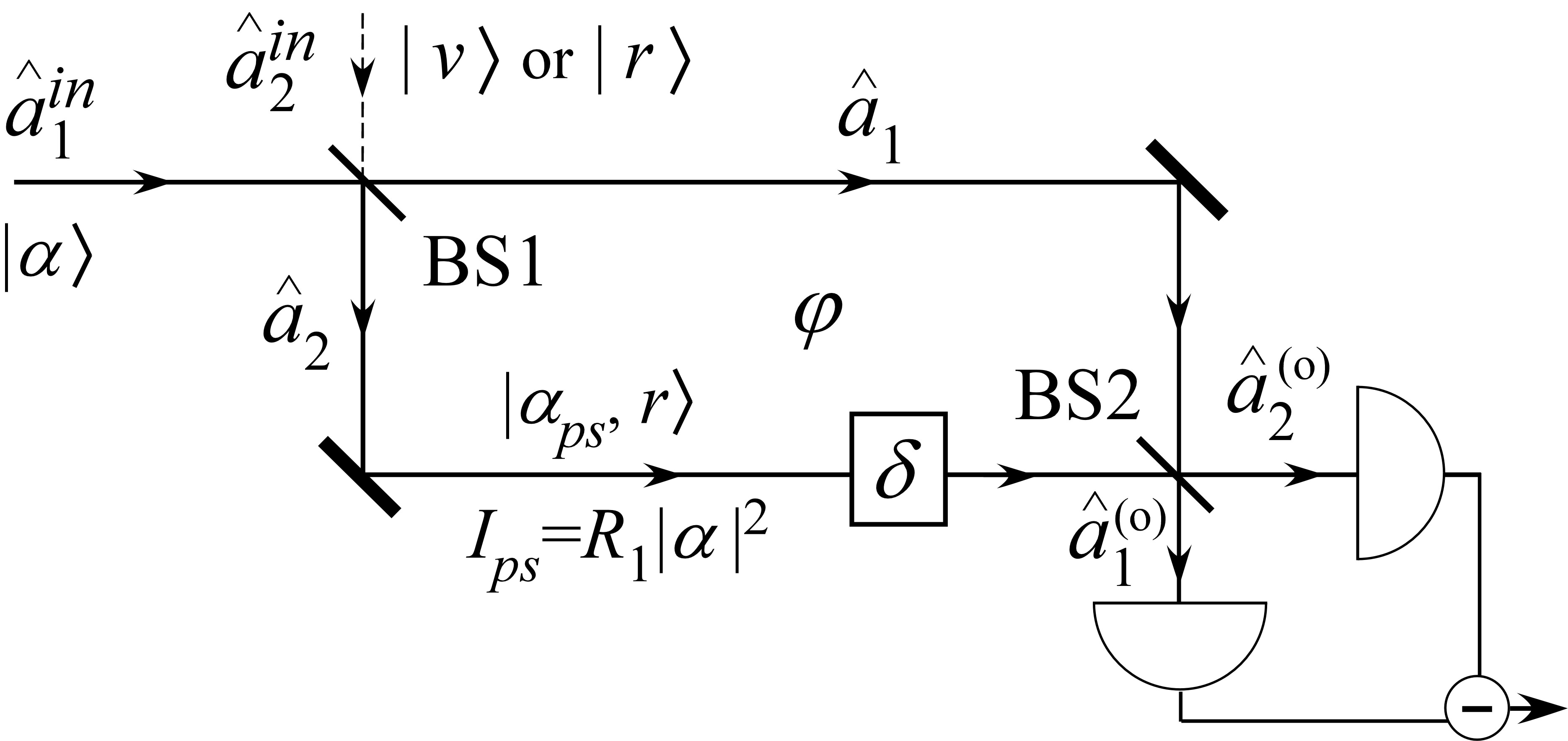}
\end{tabular}
\end{center}
\caption
{ \label{MZ2}
A classical Mach-Zehnder interferometer with vacuum ($|v\rangle$) or squeezed ($|r\rangle$) states  at the unused port (dashed line). }
\end{figure}

The interferometry technique is usually based on interferometers such as the Mach-Zehnder (MZ) type shown in Fig.\ref{MZ2}, where an incoming field in a coherent state of $|\alpha\rangle$ is split by a beam splitter (BS1) of transmissivity $T_1$ and reflectivity $R_1$ and then recombined by another of the same type (BS2) but of transmissivity $T_2$ and reflectivity $R_2$. It is straightforward to find the photon number outputs of the interferometer given by
\begin{eqnarray}
\label{eq:MZ}
I_1^{(o)} &= |\alpha|^2 (T_1T_2+ R_1R_2 - 2\sqrt{T_1T_2R_1R_2}\cos\varphi),\cr
I_2^{(o)} &= |\alpha|^2 (T_1R_2+ R_1T_2 + 2\sqrt{T_1T_2R_1R_2}\cos\varphi),
\end{eqnarray}
where $\varphi$ is the overall phase difference between the two arms of the interferometer and $T_{1,2}+R_{1,2}=1$. Note the energy conservation: $I_1^{(o)}+I_2^{(o)} = |\alpha|^2 \equiv I_{in}$. For a small phase change $\delta$, the change in the output photon number is
\begin{equation}
\label{eq:DI}
\delta I_1^{(o)} = -\delta I_2^{(o)} = 2|\alpha|^2\delta \sqrt{T_1T_2R_1R_2}  \sin\varphi.
\end{equation}
Because the two outputs are 180 degree out of phase, we can make full use of the two outputs by measuring the difference $I_-^{(o)} = I_1^{(o)}-I_2^{(o)}$, which gives twice the change:
\begin{equation}
\label{eq:DI2}
\delta I_-^{(o)} = 4|\alpha|^2 \delta\sqrt{T_1T_2R_1R_2} \sin\varphi.
\end{equation}
Obviously, the change $\delta I_-^{(o)}$ is maximum when $\varphi=\pi/2$, which we will take in the following.

The measurement sensitivity, on the other hand,  depends on the noise level at detection. For the input of a coherent state $|\alpha\rangle$, the detection noise is the photon number fluctuation, which has the Poissonian statistics: $\langle \Delta^2 I_{1,2}^{(o)}\rangle = I_{1,2}^{(o)}$. Since the two outputs are also in coherent states so that their fluctuations are uncorrelated quantum mechanically, we have
\begin{equation}
\label{eq:noise}
\langle \Delta^2 I_-^{(o)}\rangle = \langle \Delta^2 I_{1}^{(o)}\rangle+\langle \Delta^2 I_{2}^{(o)}\rangle = I_1^{(o)}+I_2^{(o)} =|\alpha|^2 .
\end{equation}
If the signal-to-noise ratio (SNR) is defined as
\begin{equation}
\label{eq:SNR}
SNR \equiv \frac{(\delta I_-^{(o)})^2}{\langle \Delta^2 I_-^{(o)}\rangle} ,
\end{equation}
we obtain the SNR for the MZ interferometer:
\begin{equation}
\label{eq:SNR2}
SNR_{MZ} = 16|\alpha|^2 \delta^2 T_1T_2R_1R_2 = 16 T_1 T_2 R_2 I_{ps} \delta^2 ,
\end{equation}
where $I_{ps} \equiv R_1|\alpha|^2 = I_2$ is the number of photons in the field that probes the phase change. This quantity is an important figure of merit for fair comparison of different schemes of phase measurement. With $T_2+R_2=1$, we find the optimum SNR:
\begin{equation}
\label{eq:SNRop}
SNR_{MZ}^{(op)} = 4 I_{ps} \delta^2
\end{equation}
when $T_2=R_2=1/2$ and  $T_1 \rightarrow 1$. The minimum measurable phase shift is $\delta_m = 1/2\sqrt{I_{ps}}$ when $SNR_{MZ}^{(op)}=1$.  This is the optimum phase measurement sensitivity that is achievable with a classical probing field for a given phase sensing photon number $I_{ps}$. Since the detection noise is from photon number fluctuation of Poissonain nature and is the same as the shot noise in an electric current, this phase measurement sensitivity is known as ``the shot noise limit (SNL)". Moreover, since the photon number fluctuation is originated from quantum nature of light, this limit is also called ``standard quantum limit (SQL)" of phase measurement.

Note that the optimum condition $T_1\approx 1$ leads to extremely unbalanced photon numbers in the two arms of the interferometer. This is in contrary to the popular balanced implementation of the interferometer \cite{xiao,ou12}, which gives rise to the controversy of two classical limits of phase measurement for comparison with quantum measurement \cite{gupta}. However, the unbalanced scheme is consistent with the homodyne measurement technique where the local oscillator (LO) has much stronger intensity than the signal field. Here in the unbalanced scheme, the phase-encoded field ($\hat a_2$) can be regarded as the signal field to be measured whereas the other arm ($\hat a_1$) with much larger photon number is treated as the LO. The condition of $T_2=R_2=1/2$ corresponds to balanced homodyne measurement \cite{yuan}. Thus, the balanced homodyne measurement technique achieves the optimum phase measurement sensitivity in classical interferometry.

Perhaps a better way to understand why we need to have an unbalanced interferometer for optimum sensitivity is through the intrinsic phase uncertainty $\Delta^2\varphi$ in any optical field \cite{ousu}. The interference method measures phase difference $\varphi=\varphi_1-\varphi_2$ so that the measurement uncertainty is $\Delta^2\varphi=\Delta^2\varphi_1+\Delta^2\varphi_2$ if the phase fluctuations in the two arms are independent (indeed, the quantum fluctuations are independent for coherent states). However, it was shown \cite{ousu} that the intrinsic phase uncertainty $\Delta^2\varphi_i (i=1,2)$ is inversely proportional to $I_i$. Making $I_1\gg I_2$ gives $\Delta^2\varphi_1\ll\Delta^2\varphi_2$ so that $\Delta^2\varphi\approx \Delta^2\varphi_2=\Delta^2\varphi_{ps}\sim 1/I_{ps}$ (the subscript $ps$ denotes the phase sensing field). But for a balanced interferometer, $I_1=I_2$ or $\Delta^2\varphi_1=\Delta^2\varphi_2$ so we have $\Delta^2\varphi=2\Delta^2\varphi_{ps}$. Hence, the unbalanced interferometer has half the measurement uncertainty as the balanced one \cite{ou12} and thus better sensitivity with twice the SNR \cite{gupta}.

The shot noise limit can be surpassed if we inject a squeezed state $|r\rangle$ into the unused port $\hat a_2^{in}$ (dashed line) of the interferometer \cite{cav}, as shown in Fig.\ref{MZ2}. Under the optimum operational condition of $T_2=R_2=1/2$ and  $T_1 \rightarrow 1, R_1\ll 1$, the probe field becomes a coherent squeezed state $|\alpha_{ps}, r\rangle$ with $\alpha_{ps}  = \alpha\sqrt{R_1}$ and squeezing parameter $r$, which is related to the amplitude gains $G = \cosh r, g=\sinh r (r>0)$ of a degenerate parametric amplifier generating the squeezed state \cite{walls}. As mentioned before, the second BS is equivalent to a balanced homodyne measurement and it is straightforward to find the photon number fluctuation at this time as \cite{yuan}
\begin{equation}
\label{eq:sq-noise}
\langle \Delta^2 I_-^{sq}\rangle =|\alpha|^2 e^{-2r} = |\alpha|^2/(G+g)^2,
\end{equation}
which gives rise to the signal-to-noise ratio as
\begin{equation}
\label{eq:sq-SNR}
SNR_{MZ}^{sq} = 4I_{ps}\delta^2 (G+g)^2.
\end{equation}
Note that this SNR for the squeezed state interferometry has an enhancement factor of $(G+g)^2$ compared to the optimum classical SNR in Eq.(\ref{eq:SNRop}). Since the detection noise in Eq.(\ref{eq:sq-noise}) is smaller than the shot noise level in Eq.(\ref{eq:noise}), this leads to the sub-shot noise interferometry \cite{xiao,gran}.

In the expressions above, we assumed $R_1|\alpha|^2 \gg g^2=\sinh^2 r$ so that the coherent state provides most of the photons for phase sensing. At large $r$-value, the squeezed state contributes a sizable photon number for $I_{ps}$ and optimization between $r$ and $\alpha$ will lead to the so-called Heisenberg limit of phase measurement \cite{bon82}.

In practice, interferometers are operated at the dark fringe mode with $\varphi = \pi$ and $T_1=T_2\gg R_1=R_2$ and homodyne measurement is performed at the dark port ($\hat a_2^{(o)}$). This is because high sensitivity requires high $I_{ps}$ (see Eq.(\ref{eq:SNRop})), which can saturate the detectors. At the dark port, the output noise is simply the vacuum noise or the squeezed noise from the unused input port ($\hat a_2^{(in)}$ in Fig.\ref{MZ2}) so it can be easily shown \cite{cav} that the SNR in this case is the same as the optimized classical SNR given by Eq.(\ref{eq:SNRop}) or the squeezed state case given by Eq.(\ref{eq:sq-SNR}).

\section{SU(1,1) interferometers}\label{sec:III}

\begin{figure}
\begin{center}
\includegraphics[width=3.2 in]{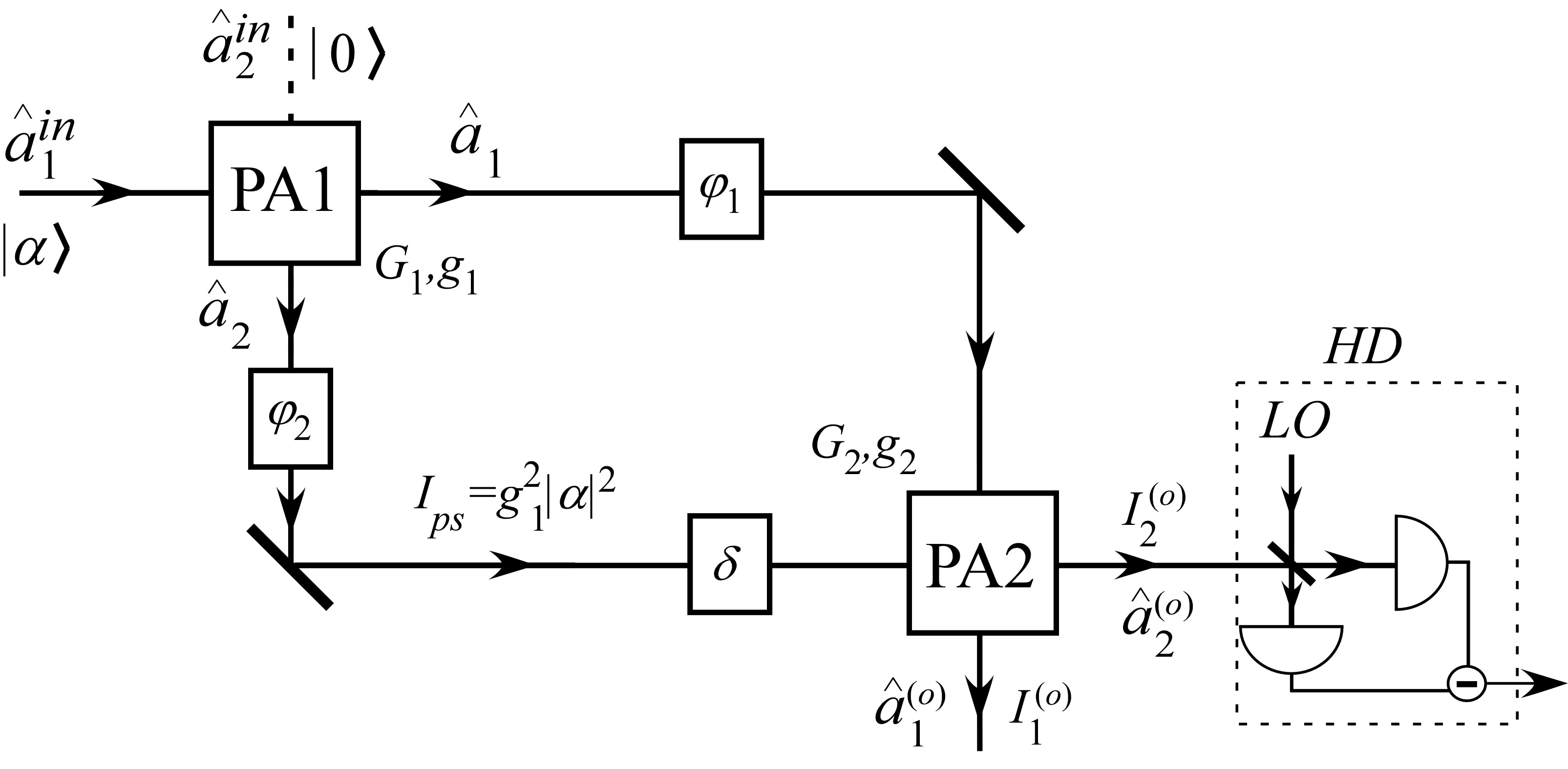}
\end{center}
\caption
{ \label{SUI} SU(1,1) interferometer, where beam splitters of traditional interferometers are replaced by parametric amplifiers (PA1, PA2) of gain $G_1,G_2$, respectively. $I_{ps}=g_1^2|\alpha|^2$ is the photon number of the field sensing the phase change $\delta$.  }
\end{figure}

A new type of interferometer, known as the so-called ``SU(1,1) interferometer"\cite{yuk,pl,ou12,jing11,hud14} is formed when we replace the beam splitters in a Mach-Zehnder interferometer with parametric amplifiers (Fig.\ref{SUI}), which can split and mix two input fields coherently for interference but possess some unique quantum behaviors. So, this new type of interferometer is of quantum nature and exhibits some advantages over the classical interferometers.

\subsection{Parametric amplifiers as beam splitters}\label{sec:IIIA}

Parametric amplifiers are a result of three-wave or four-wave mixing in a nonlinear optical process. The interaction Hamiltonian is in the form of
\begin{equation}
\label{eq:H-PA}
\hat H_{PA} = i\hbar \xi \hat a_1^{\dag} \hat a_2^{\dag} - i\hbar  \xi^*\hat a_1\hat a_2,
\end{equation}
where $\xi$ is some parameter proportional to the $\chi^{(2)}$-nonlinear coefficient and the amplitudes of strong pump fields which can be treated as classical waves. The other two relatively weak fields are the quantum fields described by the operators $\hat a_1, \hat a_2$. To have better comparison with classical interferometers, we use Heisenberg picture here and describe the system with operator evolution. The input and output relation of the quantum fields for the Hamiltonian in Eq.(\ref{eq:H-PA}) is
\begin{equation}
\label{eq:in-out}
\hat a_1^{(o)} = G \hat a_1^{(in)} + g \hat a_2^{(in)\dag},~~\hat a_2^{(o)} = G \hat a_2^{(in)} + g \hat a_1^{(in)\dag}
\end{equation}
with $G = \cosh|\eta|, g=\sinh|\eta|$ as the amplitude gains and $\eta\propto \xi$. Note that we set the phase of $\eta$ to be zero for convenience without loss of generality.

If the quadrature-phase amplitudes are defined as $\hat X=\hat a+\hat a^{\dag}, \hat Y=i(\hat a^{\dag}-\hat a)$, we have from Eq.(\ref{eq:in-out})
\begin{equation}
\label{eq:XYin-out}
\hat X_{1,2}^{(o)} = G \hat X_{1,2}^{(in)} + g \hat X_{2,1}^{(in)},~~\hat Y_{1,2}^{(o)} = G \hat Y_{1,2}^{(in)} - g \hat Y_{2,1}^{(in)}
\end{equation}
Note from the relation above that the output amplitudes are  mixtures of the two input amplitudes and thus parametric amplifiers can act as beam splitters for wave splitting and mixing. The difference is that the outputs for parametric amplifiers are amplified because $G=\cosh|\eta| > 1$.

\subsection{Interference fringe patterns}\label{sec:IIIB}

Using Eq.(\ref{eq:in-out}) for the two parametric amplifiers and assuming input field $\hat a_1^{in}$ is in a relatively strong coherent state $|\alpha\rangle (|\alpha|^2\gg1)$ and $\hat a_2^{in}$ in vacuum, and  the fields in the two arms experience phase shifts of $\varphi_{1,2}$, we find the output photon numbers as
\begin{eqnarray}
\label{eq:SUI}
I_1^{(o)} &=& |\alpha|^2 [G_1^2G_2^2+ g_1^2g_2^2 + 2G_1G_2g_1g_2\cos(\varphi_1+\varphi_2)],\cr
I_2^{(o)} &=& |\alpha|^2 [G_1^2g_2^2+ G_2^2g^2_1 + 2G_1G_2g_1g_2\cos(\varphi_1+\varphi_2)],~~~~~~~
\end{eqnarray}
where $G_1,g_1$ and $G_2,g_2$ are the amplitude gains of the two parametric amplifiers, respectively.

Comparing Eq.(\ref{eq:SUI}) to Eq.(\ref{eq:MZ}), we find three unique features that differentiate SU(1,1) interferometers (SUI) from Mach-Zehnder interferometers (MZI):

\noindent ($i$) The two outputs of SUI are in phase in contrast to 180 degree out of phase for MZI in Eq.(\ref{eq:MZ});

\noindent ($ii$) The interference fringes depend on the phase sum of $\varphi_1, \varphi_2$ instead of phase difference in MZI;

\noindent ($iii$) The outputs are amplified when the gain parameters $G_2,g_2$ are large.

The first property of in-phase fringes was demonstrated experimentally in the first realization of SU(1,1) interferometer \cite{jing11} and in the atom-light hybrid interferometer \cite{chen15a}. This property leads to $I_1^{(o)}-I_2^{(o)} =|\alpha|^2$,  which, unlike Eq.(\ref{eq:DI2}) of the classical MZI, is completely independent of the phase, making it impossible to obtain any phase change information in the intensity difference between the two outputs. This also indicates that the photon numbers of the two outputs are highly correlated, which is a property of parametric processes known as the Manley relation \cite{boyd}. The second property makes it impossible to have the common path rejection property in such devices as Sagnac interferometers but can give rise to signal enhancement when both beams are used to probe the phase change, as we will show in Sect.\ref{sec:IVC}. The third property leads to the enhancement of the signal size due to a small phase change $\delta$ on $\varphi_2$ of the probe field:
\begin{eqnarray}
\label{eq:DI-SUI}
\delta I_1^{(o)} = \delta I_2^{(o)} &=& 2\delta|\alpha|^2 G_1G_2g_1g_2  \sin(\varphi_1+\varphi_2)\cr
&\approx & 2G_2g_2 I_{ps} \delta \sin(\varphi_1+\varphi_2)~~{\rm for}~~ g_1\gg 1,~~~~~~
\end{eqnarray}
where $I_{ps} \equiv g_1^2|\alpha|^2 = I_2$. The enhancement factor is $G_2g_2$ as compared to the MZI in Eq.(\ref{eq:DI2}) at optimum condition of $R_1\ll 1, T_2=R_2=1/2$. This is because of the amplification of the second parametric amplifier when it mixes the two interfering fields.

\subsection{Quantum noise performance of SU(1,1) inteferometers}\label{sec:IIIC}

Although the signal due to phase change is increased in SUI as compared to MZI, one may argue that this is not surprising at all because of the amplification of the second PA in SUI. We can achieve the same effect if we place an amplifier at the outputs of the MZ interferometer. However, as we will see, there is a significant difference in the noise performance. An amplifier at the outputs of the MZI will amplify not only the signal but also the noise. As a matter of fact, due to added noise from its internal degrees of freedom, such an amplifier often degrades the signal-to-noise ratio, leading to reduced measurement sensitivity \cite{caves82,walls87,ou93a}. The SUI, on the contrary, will not amplify the noise as much as the signal, leading to an enhancement of the signal-to-noise ratio. The key is in the destructive interference of the quantum noise  and it can be understood from the following three perspectives.

\subsubsection{Quantum noise reduction by destructive quantum interference}\label{sec:IIIC1}

Assume a coherent state $|\alpha\rangle$ input to the SUI and we make homodyne detection of $\hat X(\theta)=\hat a e^{-i\theta}+\hat a^{\dag} e^{i\theta}$ at the outputs of PA2. It is straightforward to calculate the quantum fluctuations as \cite{ou12}
\begin{eqnarray}
\label{eq:SUI-noise}
&&\langle \Delta^2 \hat X_1^{(o)}(\theta)\rangle  = \langle \Delta^2 \hat X_2^{(o)}(\theta)\rangle \cr && \hskip 0.15in = |G_1G_2+g_1g_2e^{i(\varphi_1+\varphi_2)}|^2+|G_1g_2+g_1G_2e^{i(\varphi_1+\varphi_2)}|^2\cr  && \hskip 0.15 in =
(G_1^2+g_1^2)(G_2^2+g_2^2) + 4G_1G_2g_1g_2\cos(\varphi_1+\varphi_2).~~~~~~
\end{eqnarray}
The dependence on $\varphi_1+\varphi_2$ is a result of quantum interference in SUI, just as in the output photon numbers in Eq.(\ref{eq:SUI}).  Although we find from Eq.(\ref{eq:DI-SUI}) that the measured signal is maximum when $\varphi_1+\varphi_2=\pi/2$,  the minimum noise is achieved at the dark fringe when $\varphi_1+\varphi_2=\pi$ and at balanced gain of $G_2=G_1, g_2=g_1$ for a given $G_1,g_1$:
\begin{eqnarray}
\label{eq:SUI-noise2}
\langle \Delta^2 \hat X_1^{(o)}(\theta)\rangle_m & =& \langle \Delta^2 \hat X_2^{(o)}(\theta)\rangle_m \cr
&=& 1 +2(G_1g_2-G_2g_1)^2 \cr &=& 1~~{\rm when}~~ G_1=G_2, g_1=g_2.
\end{eqnarray}
Since the noise in each arm of the SUI after PA1 is $G_1^2+g_1^2 =1+2g_1^2 >1$, the noise is reduced at the outputs of the SUI (PA2). This is because of the destructive quantum interference between the two arms that cancels the large quantum noise at each arm. Such a noise reduction effect was observed by Hudelist {\it et al.} in the first measurement of quantum noise performance of SUI \cite{hud14}.

Note that Eq.(\ref{eq:SUI-noise2}) is independent of angle $\theta$, which means that the noise is minimum for all quadrature-phase amplitudes $\hat X_{1,2}^{(o)}(\theta)$. This is quite different from squeezed state interferometry\cite{cav,xiao,gran} where only the squeezing quadrature has noise reduction. This indicates that the underlying physics for noise reduction here is quantum destructive interference, which reduces noise for the whole field including all quadrature-phase amplitudes, in contrast to the squeezed state interferometry where noise depends on the angle of quadrature-phase amplitudes.

\subsubsection{Quantum beam splitter as a disentanglement tool}\label{sec:IIIC2}

To understand how quantum interference occurs at PA2, we just need to recall Eq.(\ref{eq:XYin-out}), which shows the superposition of the quadrature-phase amplitudes of the incoming fields. Note that the relations are in quantum mechanical operator form, which means that quantum fluctuations or noise can be subtracted or added depending on the phase, giving rise to quantum interference. This shows that a parametric amplifier can act as a quantum beam splitter to split and mix waves. In this sense, the roles of a PA and a BS are the same in the mixing of waves: incoming waves are all superposed coherently. { It is known that the two outputs of PA1 are entangled in the continuous variables of phases and amplitudes \cite{reid,ou92} and two entangled fields can be transformed into two independent squeezed states with noise reduced at orthogonal quadratures \cite{ou92b,brau,kong13}. In this case, the BS acts as a disentangler that transforms two entangled fields into two unentangled fields. Since a PA and a BS are the same in wave mixing, the role of PA2 in the SU(1,1) interferometer is then a disentangler, producing two unentangled fields at the outputs. }

On the other hand, the difference between a parametric amplifier (PA) and a linear beam splitter (BS) lies in the fact that a parametric amplifier (PA2) has amplified outputs. This feature can lead to loss-tolerant property of SU(1,1) interferometers that we will discuss later in Sect.\ref{sec:IIIE}. It also leads to the following understanding.

\subsubsection{Quantum noiseless amplification due to noise cancelation}\label{sec:IIIC3}

The action of the SU(1,1) interferometer can be analyzed from another perspective, that is, quantum amplification. When viewed as an amplifier, one of the inputs of PA2 is regarded as the signal input while the other input is treated as the internal mode of the amplifier \cite{caves82,walls87,ou93a}. Normally, the internal mode of the amplifier is inaccessible from outside and is left in vacuum state, which adds in vacuum noise to the amplified signal. This added noise is the extra noise in addition to the input signal noise, leading to degraded signal-to-noise ratio for the amplified signal compared to the input. If the internal mode can be accessed, as in the case of a parametric amplifier, squeezed states can be injected to it to reduce the extra added noise \cite{walls87,ou93a}. This is the case when the input signal and the internal mode are uncorrelated.
On the other hand,  if the input signal and the internal mode are correlated, further noise reduction can be achieved. This was first studied by Ou \cite{ou93} as early as in 1994 and recently was demonstrated \cite{kong13b} with an arrangement similar to an SU(1,1) interferometer. To understand this, we go back to the input-output relation in Eq.(\ref{eq:XYin-out}) for the parametric amplifier. We select field 1 as the signal field (s) and field 2 as the internal mode (int) and rewrite it as
\begin{equation}
\label{eq:Ampin-out}
\hat X_{s}^{(o)} = G \hat X_{s}^{(in)} + g \hat X_{int}^{(in)}.
\end{equation}
If the signal and internal mode are independent, we have
\begin{equation}
\label{eq:Amp-un}
\langle \Delta^2 \hat X_{s}^{(o)}\rangle  = G^2 \langle \Delta^2 \hat X_{s}^{(in)}\rangle + g^2 \langle \Delta^2 \hat X_{int}^{(in)}\rangle.
\end{equation}
The second term in the expression above is the extra noise for the output that degrades the output SNR as compared to the input. But if the input signal and the internal mode are correlated, we have from Eq.(\ref{eq:Ampin-out})
\begin{equation}
\label{eq:Amp-co}
\langle \Delta^2 \hat X_{s}^{(o)}\rangle  = G^2 \langle \Delta^2 (\hat X_{s}^{(in)} + \lambda \hat X_{int}^{(in)})\rangle
\end{equation}
with $\lambda \equiv g/G$. If the signal and the internal modes are in the EPR-type entangled state such as those generated from the first PA, $\hat X_{s}^{(in)}$ and $\hat X_{int}^{(in)}$ are quantum mechanically correlated so that $\langle \Delta^2 (\hat X_{s}^{(in)}+ \lambda \hat X_{int}^{(in)})\rangle$ can be smaller than the corresponding value of $1+\lambda^2$ when they are both in vacuum. In fact, it was shown \cite{ou93} that noiseless quantum amplification can be achieved with the proper adjustment of the parameter. Such an effect of noise reduction in amplifier due to entanglement was  demonstrated first by Kong {\it et al.}\cite{kong13b} in atomic vapor system and later by Guo {\it et al.}\cite{guo16} in nonlinear fiber amplifier.

\subsection{Signal-to-noise ratio and the optimum phase measurement sensitivity in SU(1,1) interferometer}\label{sec:IIID}

\begin{figure}
\begin{center}
\includegraphics[width=3.2 in]{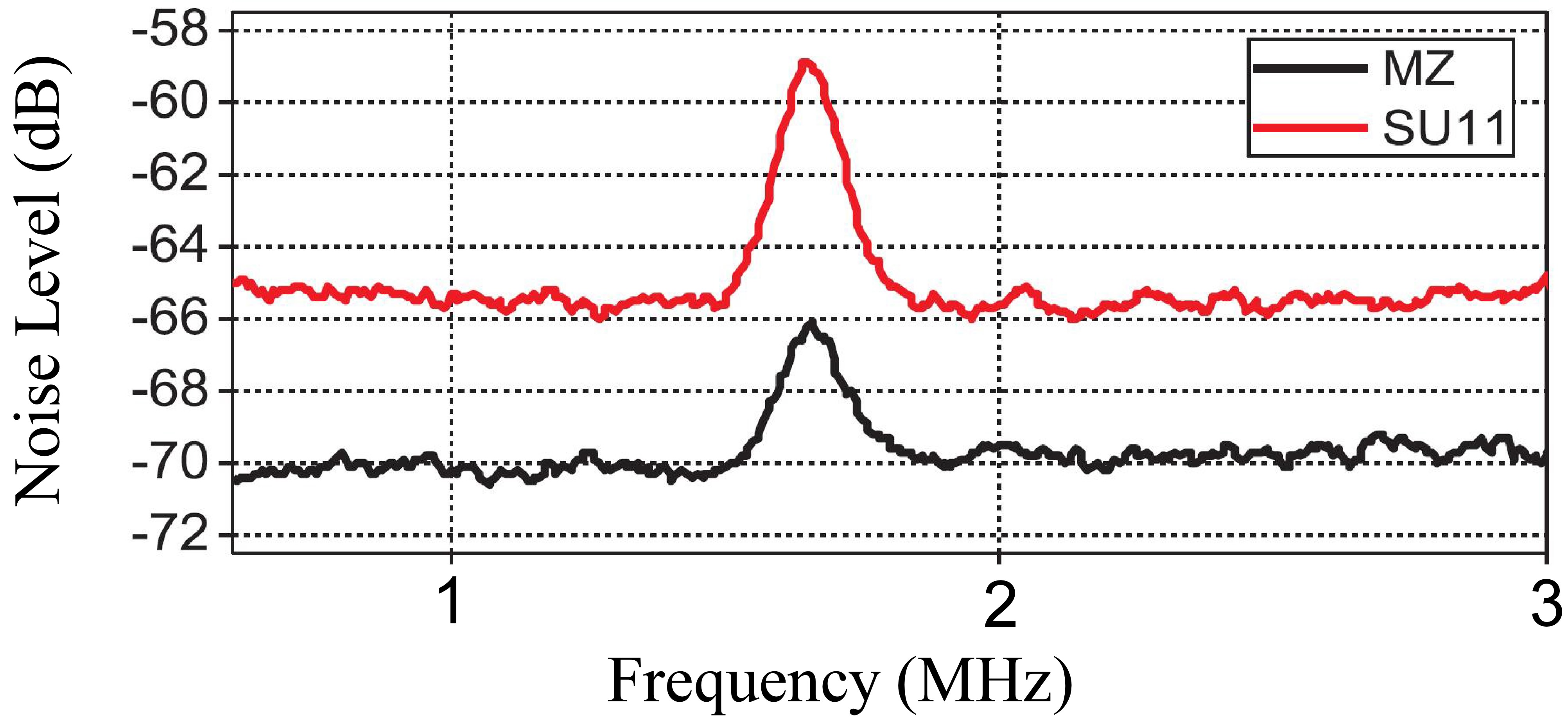}
\end{center}
\caption
{ \label{fig:SNR}
Phase modulation signals and noise levels for an SU(1,1) interferometer (red) and a Mach Zehnder interferometer (black). Adapted from Ref.\citenum{du18}. }
\end{figure}

The sensitivity of SU(1,1) interferometer for phase measurement is determined not only by the noise level of the outputs but also by the signal size due to phase change. It is usually characterized by the signal-to-noise ratio (SNR).  Although the signal size due to phase change is usually the best at half of the fringe size, i.e., the overall phase is at $\pi/2$,  it is better to operate at the dark fringe for practical reasons, similar to the Mach-Zehnder interferometer in Sect.\ref{sec:II}, and we make homodyne measurement of $\hat Y=i(\hat a^{\dag}-\hat a)$. Referring to Fig.\ref{SUI}, when the input field 1 to the interferometer is in a coherent state of $|\alpha\rangle$ and the overall phase $\varphi_1+\varphi_2$ is set at $\pi$ for minimum at both outputs, we obtain the signals at the two outputs for a small phase change $\delta$ in one arm of the interferometer \cite{JML18}:
\begin{equation}
\label{eq:signal}
\langle \hat Y_{1}^{(o)}\rangle  = 2g_1g_2|\alpha|\delta, ~~~~\langle \hat Y_{2}^{(o)}\rangle  = 2g_1G_2|\alpha|\delta
\end{equation}
With the output noise given in Eq.(\ref{eq:SUI-noise}), we obtain the SNRs at the two outputs as
\begin{eqnarray}
\label{eq:SUI-SNR-1}
SNR_{SUI}^{(1)} &=& \frac{\langle \hat Y_{1}^{(o)}\rangle^2}{\langle \Delta^2\hat Y_{1}^{(o)}\rangle}\cr
& =&  \frac{4g_2^2g_1^2|\alpha|^2\delta^2}{(G_1^2+g_1^2)(G_2^2+g_2^2) -4G_1G_2g_1g_2}\cr
&=& \frac{4g_2^2I_{ps}\delta^2}{(G_1^2+g_1^2)(G_2^2+g_2^2) -4G_1G_2g_1g_2},
\end{eqnarray}
and
\begin{eqnarray}
\label{eq:SUI-SNR-2}
SNR_{SUI}^{(2)}
& =&  \frac{4G_2^2g_1^2|\alpha|^2\delta^2}{(G_1^2+g_1^2)(G_2^2+g_2^2) -4G_1G_2g_1g_2}\cr
&=& \frac{4G_2^2I_{ps}\delta^2}{(G_1^2+g_1^2)(G_2^2+g_2^2) -4G_1G_2g_1g_2},
\end{eqnarray}
where $I_{ps}=g_1^2|\alpha|^2$ is the photon number of the phase sensing field. When $g_2\rightarrow \infty$ so that $G_2^2=1+g_2^2\approx g_2^2$, the SNR takes the maximum value of
\begin{equation}
\label{eq:SUI-SNR2}
SNR_{SUI}^{(1,2) op} = 2(G_1+g_1)^2I_{ps}\delta^2.
\end{equation}
Notice that the optimum SNR is obtained not with equal gains of the two PAs but under the condition of $G_2 \gg 1$.\cite{manc17}
Figure \ref{fig:SNR} shows a typical result of phase measurement, from which the SNR can be extracted,  by both an SU(1,1) interferometer (red) and a Mach-Zehnder interferometer (black) under the condition of the same phase sensing intensity \cite{du18}. The peaks are from phase modulation signal and the flat floor is the noise level of measurement. Since it is in log-scale, the SNR of phase measurement is simply the difference of the peak value and the floor value. It is found that $SNR_{SUI} = 6.9 dB$ and $SNR_{MZI} = 3.9 dB$, leading to an improvement of $3.0 dB$ in SNR by SUI over MZI. Notice that both the signal and noise of SUI are amplified from MZI but with the signal amplified more.

Since both outputs contain the information about the phase change, it is suggested \cite{and17,gup18} to measure the joint quantity $\hat Y_{JM}\equiv \hat Y_1^{(o)}+\hat Y_2^{(o)}$ to combine the information. It is straightforward to show\cite{JML18} in this case, the SNR is independent of $g_2$ and has the optimum value given in Eq.(\ref{eq:SUI-SNR2}): $SNR_{JM} = SNR_{SUI}^{(1,2) op} = 2(G_1+g_1)^2I_{ps}\delta^2$ but with no further improvement.

Comparing Eq.(\ref{eq:SUI-SNR2}) to the optimum classical SNR in Eq.(\ref{eq:SNRop}), we obtain an SNR enhancement factor of $(G_1+g_1)^2/2$. This is a factor of 2 smaller than that of the squeezed state interferometry given in Eq.(\ref{eq:sq-SNR}). The reason for this is related to the optimum scheme of SUI for phase measurement and will be discussed later in Sect.\ref{sec:IVC}.

\subsection{Effect of losses}\label{sec:IIIE}

It is well-known that with some loss $L$ such as detection inefficiency involved in the squeezed state, the noise reduction effect is degraded with Eq.(\ref{eq:sq-noise}) modified to
\begin{equation}
\label{eq:sq-noise2}
\langle \Delta^2 I_-^{sq}\rangle =|\alpha|^2 \big[(1-L)e^{-2r}+L\big],
\end{equation}
where the loss $L$ can be modeled as a beam splitter with a transmissivity of $1-L$, and the above can be considered as contributions from two parts: the transmitted squeezed noise $|\alpha|^2e^{-2r}$ with a probability $1-L$ and the vacuum noise of size $|\alpha|^2$ reflected from the unused port with a probability of $L$, all scaled to the shot noise level of $\langle \Delta^2 I_-^{snl}\rangle = |\alpha|^2$. Notice that in the existence of loss $L$, the best noise reduction achievable is $L$, even with infinite squeezing ($r\rightarrow \infty$). After considering the loss of the signal due to loss, we arrive at the best SNR enhancement factor as $(1-L)/L$ for the squeezed state interferometry.

\begin{figure}
\begin{center}
\includegraphics[width=3.0 in]{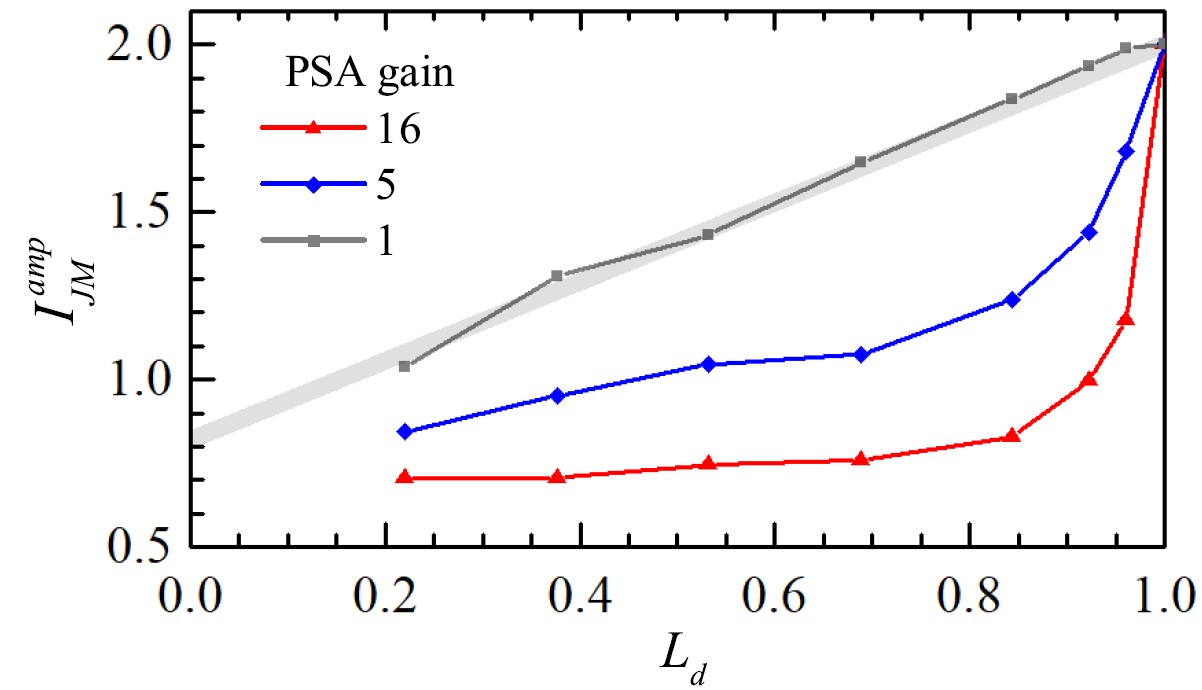}
\end{center}
\caption
{ \label{loss}
Dependence of measured quantum noise level as a function for the detection losses for various gain of parametric amplifier (PSA gain). The value of 2 corresponds to vacuum noise level. Reproduced from Li {\it et al.}, Opt. Express {\bf 27}, 30552 (2019). }
\end{figure}

On the other hand, the output noise for the SUI is amplified by the second PA, making it much larger than the vacuum noise level so that the extra noise coupled in through loss is negligible. This is shown in Eq.(\ref{eq:SUI-noise}), which becomes
\begin{eqnarray}
\label{eq:SUI-noise3}
\langle \Delta^2 \hat X_1^{(o)}(\theta)\rangle & =& \langle \Delta^2 \hat X_2^{(o)}(\theta)\rangle \cr &=&
(G_1^2+g_1^2)(G_2^2+g_2^2) - 4G_1G_2g_1g_2 \cr &=& 1 +2(G_1g_2-G_2g_1)^2
\cr &\gg &1 ~~{\rm for}~~G_2\gg G_1>1
\end{eqnarray}
at dark fringe when $\varphi_1+\varphi_2=\pi$. So, the noise for $\hat Y_1^{(o)}$ after the loss $L$ is
\begin{eqnarray}
\label{eq:SUI-noise4}
\langle \Delta^2 \hat Y_1^{(o)}\rangle_L &=& (1-L) \langle \Delta^2 \hat Y_1^{(o)}\rangle +L \cr &\approx &(1-L) \langle \Delta^2 \hat Y_1^{(o)}\rangle.
\end{eqnarray}
With the signal drop by a factor of $1-L$: $\langle \hat Y_{1}^{(o)}\rangle_L^2= (1-L)\langle \hat Y_{1}^{(o)}\rangle^2$, we obtain the SNR due to loss:
\begin{eqnarray}
\label{eq:SUI-SNR-L}
SNR_{SUI}^{L} = \frac{\langle \hat Y_{1}^{(o)}\rangle_L^2}{\langle \Delta^2\hat Y_{1}^{(o)}\rangle_L}\approx
\frac{\langle \hat Y_{1}^{(o)}\rangle^2}{\langle \Delta^2\hat Y_{1}^{(o)}\rangle} = SNR_{SUI}.
\end{eqnarray}
So, the losses outside of the interferometer such as transmission and detection losses have almost no effect on the SNR of SUI for large $G_2$ and the ability of loss-tolerance increases with $G_2$ of PA2 \cite{ou12,JML18}.  This loss-tolerant property of SUI was first observed in Ref.\citenum{hud14} and confirmed later in Refs.\citenum{che17} and \citenum{li19}. Figure \ref{loss} shows the result from Ref.\citenum{li19}, which plots the measured quantum noise level (value of 2 corresponds to vacuum level) as a function of loss for various gain of parametric amplifier. It clearly demonstrates that the effect of loss is mitigated by the amplification. The straight gray line corresponds to the case of direct detection and is described by the linear dependence in Eq.(\ref{eq:sq-noise2}).

In fact, the amplified quantum noise from PA2 can not only overcome the vacuum noise introduced through losses, it can also fight against excess classical noise. This strategy was used in microwave detection to tackle the enormous thermal noise background in microwave circuit \cite{fl12} (see Sect.\ref{sec:VA}).

SUI's immunity to losses is only for the output fields of the SUI. For losses inside the interferometer, however, it was shown \cite{ou12,mar12} that the effect is exactly the same as that on the squeezed state. So, SUI is not immune to its internal losses. This suggests that all the quantum advantage is from the quantum entanglement created in the first PA (PA1) whereas the second PA is simply a device for superposition to disentangle the two fields in the two arms of the interferometer.

Indeed, as variations of SUI, we can replace the second PA with any linear device that can mix the two fields and achieve the same performance as SUI, as we will see in the following.

\section{Variations of SU(1,1) interferometers}\label{sec:IV}

\subsection{The Scheme of a parametric amplifier and a beam splitter (PA+BS)}\label{sec:IVA}

It has been known almost since the discovery of squeezed states and EPR entangled states that in the case of degenerate frequency, they can be converted from each other by a 50:50 beam splitter \cite{ou92b,fur99}. Since an EPR-type entangled state can be generated by a parametric amplifier \cite{ou92}, we can use a beam splitter to convert it to squeezed states and measure the phase change with reduced quantum noise, similar to the squeezed state interferometry. However, the statements above are for states no coherent components but the photon number of squeezed states with no coherent component is too low to have any practical use.

\begin{figure}
\begin{center}
\includegraphics[width=3.2 in]{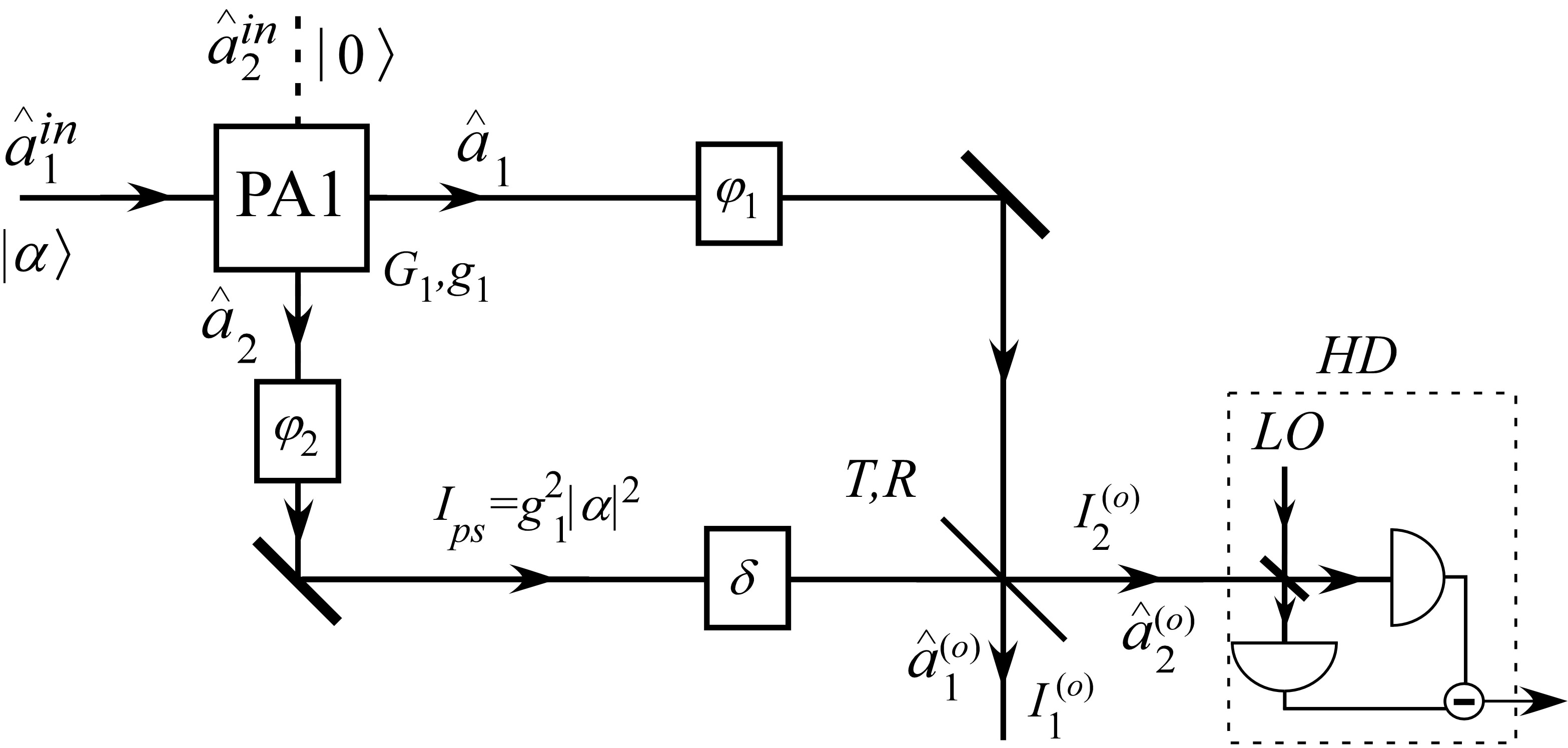}
\end{center}
\caption
{ \label{PA+BS}
The scheme of parametric amplifier and beam splitter for a variation of the SU(1,1) interferometer. Adapted from Kong {\it et al.}, Phys. Rev. A {\bf 87}, 023825 (2013). }
\end{figure}

To boost the photon number, we can inject a coherent state, just like what we did in Sect.\ref{sec:III}. This forms a variation of the SU(1,1) interferometer with a PA for beam splitting and a BS for wave superposition and interference (PA+BS scheme). The actual scheme is shown in Fig.\ref{PA+BS}. For a large injection $|\alpha|^2\gg 1$, it is straightforward to calculate \cite{kong13} the output intensity at output port 2 as
\begin{eqnarray}
\label{eq:SUI-I-PABS}
I_2^{(o)} = I_{ps} [1-{\cal V}\cos(\varphi_1+\varphi_2)],
\end{eqnarray}
where $I_{ps}= g_1^2|\alpha|^2$ and visibility ${\cal V} \equiv 2G_1g_1\sqrt{TR}/(g_1^2+R)$. Note that the fringe depends on the sum of the phases of the two arms, similar to Eq.(\ref{eq:SUI}). 100\% visibility in interference fringe at output port 2 can be achieved with $T=G_1^2/(G_1^2+g_1^2), R=1-T$ for the beam splitter. However, when $\hat Y_2^{(o)}$ is measured at output port 2 by homodyne detection (HD), the optimum SNR for phase measurement is achieved when $T=(G_1^2+g_1^2)^2/(8G_1^2g_1^2+1), R=4G_1^2g_1^2/(8G_1^2g_1^2+1)$ with
\begin{eqnarray}
\label{eq:SUI-SNR-PABS}
SNR_{PA-BS}^{(op)} = 4\delta^2 I_{ps}(G_1^2+g_1^2).
\end{eqnarray}
This is a factor of $G_1^2+g_1^2$ improvement over the optimum classical SNR in Eq.(\ref{eq:SNRop}).

If we use 50:50 beam splitter, as in Ref.\citenum{ou92b}, it is straightforward to find that the output noise will be $(G_1-g_1)^2$ while the signal is $2I_{ps}\delta^2$ and the SNR is exactly same as that in Eq.(\ref{eq:SUI-SNR2}). So, this variation of SUI gives the same SNR improvement factor as the SUI over the classical interferometer. It is interesting to note if we switch the positions of PA and BS, that is, using BS for beam splitting and PA for wave superposition, the result won't be that given in Eq.(\ref{eq:SUI-SNR-PABS}) but is the same as  that in Eq.(\ref{eq:SNRop}) for a classical interferometer \cite{kong13}. This further demonstrates that the quantum advantage originates from the quantum entanglement in the phase probing beam produced by the first parametric amplifier. Note further that since we use a BS to superpose the signal and idler fields, they must be frequency degenerate and the scheme is sensitive to losses just like squeezed state interferometry.

\subsection{Truncated SU(1,1) interferometer}\label{sec:IVB}

\begin{figure}
\begin{center}
\includegraphics[height=3.5cm]{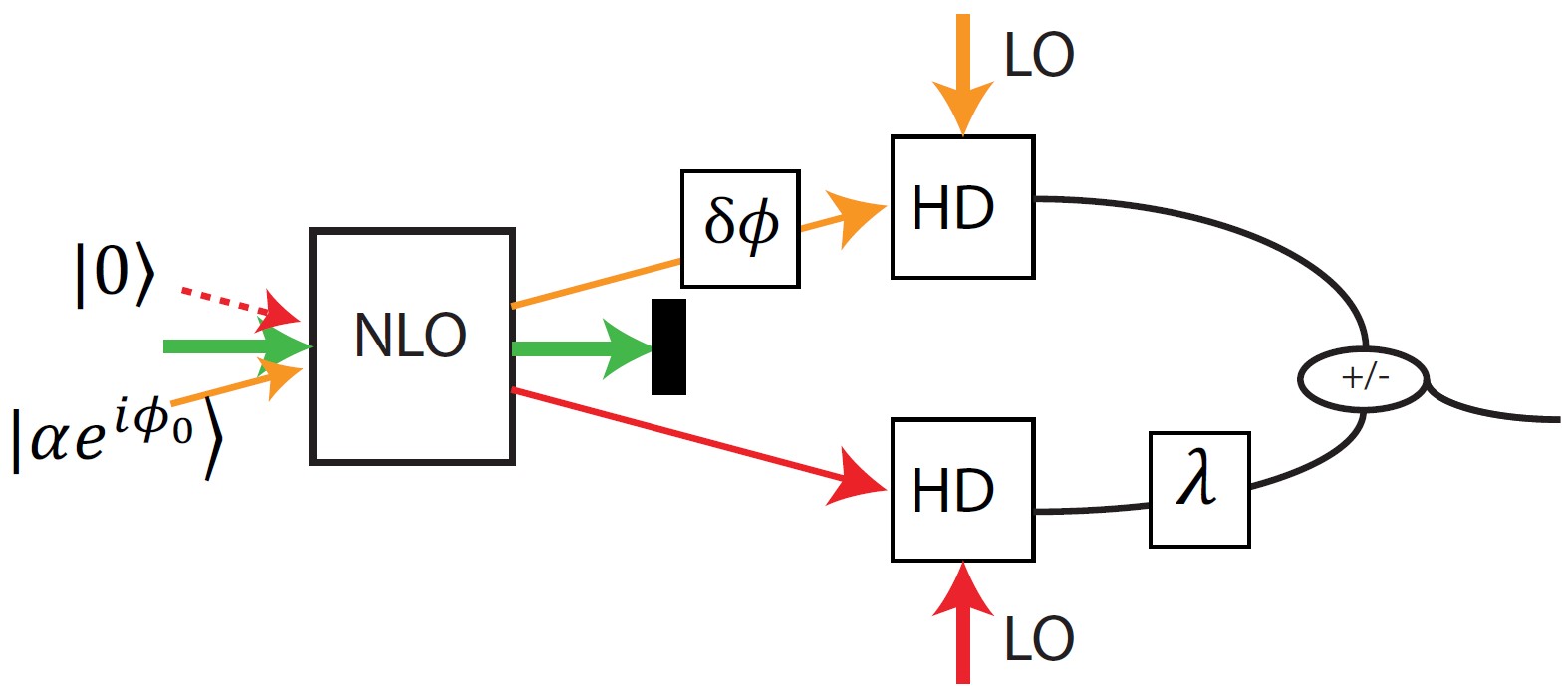}
\end{center}
\caption
{
The scheme of truncated SU(1,1) interferometer. Reproduced from Gupta {\it et al.}, Opt. Exp. {\bf 26}, 391 (2018). } \label{trun-SUI}
\end{figure}

Although waves need to be superimposed in order to show the interference effect, the method of superposition can vary. We have already seen the methods by a parametric amplifier and by a beam splitter. In these cases, the waves are physically superimposed and interference occurs at the optical fields of the outputs of the wave-combining devices. In particular for the PA+BS scheme in Fig.\ref{PA+BS}, it requires the two fields from PA1 have the same frequency because of the use of beam splitter for wave superposition. On the other hand, since homodyne detection makes quantum measurement of the quadrature-phase amplitude of the field, the photo-current from homodyne detection can be thought of as the quantum copy of the amplitude of the field. So, the mixing of the photo-currents after homodyne detections is equivalent to the superposition of the detected fields and we can replace the beam splitter with a post-detection current mixer to achieve field superposition. This is the idea behind the so-called ``truncated" SU(1,1) interferometer proposed and reported by Anderson {\it et al.}\cite{and17,gup18}, as shown in Fig.\ref{trun-SUI} where only the first parametric amplifier remains as compared to the SU(1,1) interferometers in Fig.1(b) and Fig.\ref{PA+BS}. The mixer for photo-currents from the homodyne detectors (HD) plays the same role as the second parametric amplifier in Fig.1(b) and the beam splitter in Fig.\ref{PA+BS} to superimpose the two fields in the interferometer for interference. The current after mixing shows the phase signal $\delta\phi$ as well as the quantum noise cancelation effect due to entanglement in a typical SU(1,1) interferometer.  It was shown \cite{gup18,JML18} that the SNR for phase measurement is the same as that in Eq.(\ref{eq:SUI-SNR2}) in the ideal lossless condition.

Because direct detection is involved in the truncated scheme and the PA+BS scheme of SU(1,1) interferometers, losses will have a significant effect on the quantum enhancement factor in a similar way to the squeezed state interferometry.

\subsection{Dual-beam SU(1,1) interferometers}\label{sec:IVC}

\begin{figure}
\begin{center}
\includegraphics[width=\linewidth]{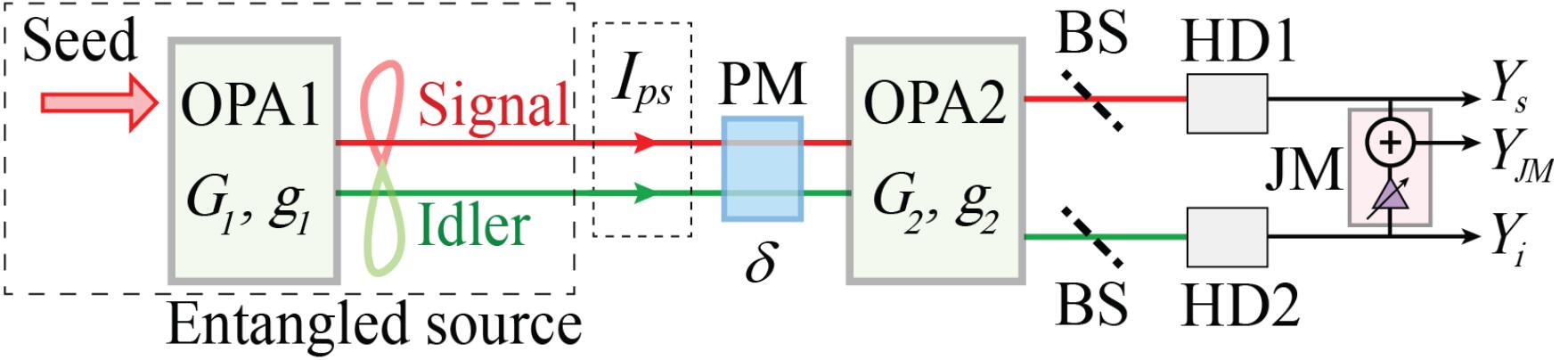}
\end{center}
\caption
{ \label{fig:dual}
The dual-beam scheme of SU(1,1) interferometer for phase measurement. Reproduced from Ref.\citenum{liu18}. }
\end{figure}

In the SU(1,1) interferometers we discussed so far,  the SNR for phase measurement is given in Eq.(\ref{eq:SUI-SNR2}), which is an improvement factor of $(G_1+g_1)^2/2$ over the optimum classical SNR in Eq.(\ref{eq:SNRop}). This is a factor of 2 smaller than the improvement factor by squeezed state interferometry given in Eq.(\ref{eq:sq-SNR}). The reason for this is quantum resource sharing in phase and amplitude measurement, which will be discussed later in Sect.\ref{sec:VIB}. This means that the current SU(1,1) interferometer is not optimized for phase measurement. To look for the optimized phase measurement scheme, we notice in Eqs.(\ref{eq:SUI}) and (\ref{eq:SUI-I-PABS}) that the interference fringe depends on the sum of the phases of the two arms of the interferometer. Therefore, if we use both fields from PA1 to sense the phase change signal, we will double the signal size $\delta$. This is the dual-beam scheme proposed by Li {\it et al.}\cite{JML18} and realized by Liu {\it et al.}\cite{liu18}, which is shown in Fig.\ref{fig:dual}. As expected, it can be shown \cite{JML18} that the homodyne detection signals at both output ports are
\begin{eqnarray}
\label{eq:signal-dual}
\langle \hat Y_{1}^{(o)}\rangle^2  &=& 4(G_1G_2+g_1g_2)^2|\alpha|^2\delta^2,\cr \langle \hat Y_{2}^{(o)}\rangle^2  &= & 4(G_1g_2+g_1G_2)^2|\alpha|^2\delta^2.
\end{eqnarray}
With the noise power given in Eq.(\ref{eq:SUI-noise}) and at dark fringe of $\varphi_1+\varphi_2=\pi$, the SNR for the dual-beam scheme is
\begin{eqnarray}
\label{eq:SNR-dual}
SNR_{DB}^{(1)}
& =&  \frac{4(G_1G_2+g_1g_2)^2I_{ps}\delta^2}{(G_1^2+g_1^2)[(G_1^2+g_1^2)(G_2^2+g_2^2) -4G_1G_2g_1g_2]},\cr
SNR_{DB}^{(2)}
& =&  \frac{4(G_1g_2+g_1G_2)^2I_{ps}\delta^2}{(G_1^2+g_1^2)[(G_1^2+g_1^2)(G_2^2+g_2^2) -4G_1G_2g_1g_2]},~~~~~~~~
\end{eqnarray}
where $I_{ps}=(G_1^2+g_1^2)|\alpha|^2$ is the photon number of the dual phase sensing fields. When $g_2\rightarrow \infty$ and $G_2\approx g_2$, we have the optimum SNR:
\begin{eqnarray}
\label{eq:SNR-dual1}
SNR_{DB}^{(1)} = SNR_{DB}^{(2)} &=& 2(G_1+g_1)^4I_{ps}\delta^2/(G_1^2+g_1^2)
\cr &\rightarrow & 4(G_1+g_1)^2I_{ps}\delta^2~~ {\rm for}~~ g_1\gg 1,~~~~~~
\end{eqnarray}
which is the same as the one for squeezed state interferometry in Eq.(\ref{eq:sq-SNR}) at large $g_1$.

Note that at finite $g_1$, the SNR in Eq.(\ref{eq:SNR-dual}) is still smaller than that for squeezed state interferometry. This is again because of quantum resource distribution, which we will discuss in Sect.\ref{sec:VIB}.  Note that the SNR in Eq.(\ref{eq:SNR-dual}) is for one output only but we have two outputs for PA2. So, we can make full use of these two outputs by performing a joint measurement $\hat Y_{JM}\equiv \hat Y_1+\lambda \hat Y_2$ of the two outputs, as shown in Fig.\ref{fig:dual}. With $\lambda=1$, it is shown that the SNR for the joint measurement is
\begin{eqnarray}
\label{eq:SNR-dual-JM}
SNR_{DB}^{JM} = 4(G_1+g_1)^2I_{ps}\delta^2~~~~~~ {\rm for~~arbitrary}~~ g_1.
\end{eqnarray}
So, we recover the SNR of squeezed state interferometry when we make full use of the resource.

It was shown \cite{JML18} that when dual-beam phase sensing is implemented in the truncated scheme and the PA+BS scheme, the factor of 2 is also recovered, leading to the same SNR as the squeezed state interferometry. But because of the second PA, the dual-beam SUI scheme here is tolerant to losses outside of the interferometer, similar to the original SU(1,1) interferometer in Fig.\ref{SUI}. Furthermore, different from the PA+BS scheme,  the employment of separate homodyne detectors in the truncated scheme and the second PA in the dual-beam SUI scheme does not require the same frequency for the two fields from the first PA in both schemes.
The experimental implementation of the dual beam SU(1,1) interferometer was realized by Liu {\it et al.}\cite{liu18} and about 3 dB improvement over the single-beam scheme was demonstrated.

\subsection{Multi-stage SU(1,1) interferometers}\label{sec:IVD}

Similar to multi-path interferometers such as Fabry-Perot interferometers and multi-slit interference in optics, we can also also add more PAs to form multi-stage SU(1,1) interferometers. In order to have all the PAs playing the same role in the multi-path interference, we usually work at low gain regime of the PAs so that spontaneous emission dominates and two-photon states are generated. This variation of SU(1,1) interferometer finds its application in the modification of mode structures (temporal and spatial) in the output field for mode engineering of the output quantum states. The detail of this application can be found later in Sect.\ref{sec:VIE}. In the following, we will present the general principle for this scheme.

\begin{figure}
\begin{center}
\includegraphics[width=3.2 in]{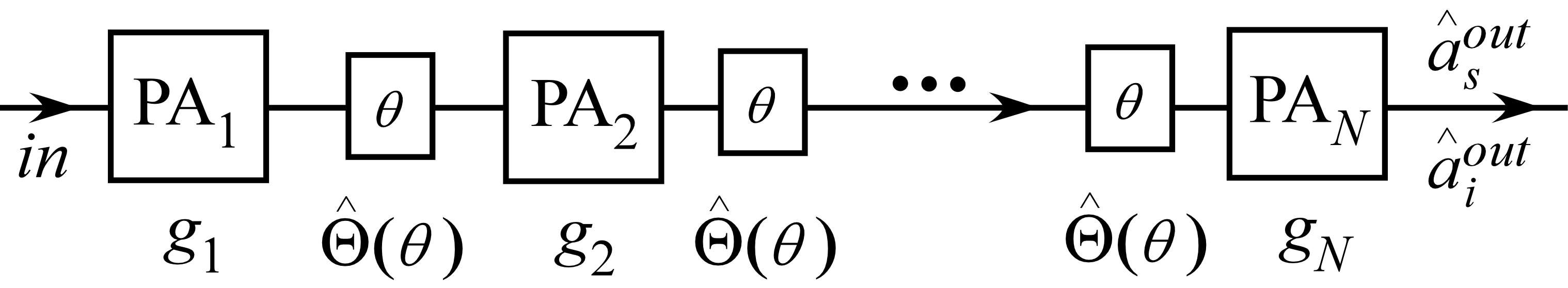}
\end{center}
\caption
{Multi-stage SU(1,1) interferometer.  }\label{fig:multi-PA}
\end{figure}

Consider the multi-stage interferometer shown in Fig.\ref{fig:multi-PA} where the $k$-th PA is described by the small amplitude gain parameter $0<g_k\ll 1$ so that the power gain $G_k^2=1+g_k^2 \approx 1 (k=1,2,...,N)$. In between the PAs, sandwiched are phase shifters $\hat \Theta(\theta)$. For simplicity, we assume the phase shifters have the same phase of the amount $\theta$ for the two fields of the PAs together. In the low gain limit, in order to better describe the performance of the system and reveal the underlying physical principle, we will work in Sch\"odinger picture with quantum states. Let's start with the quantum state of one PA.

With the Hamiltonian in Eq.(\ref{eq:H-PA}) for parametric amplifier, the state evolution in the time interval $\Delta t$ for the system is described by a unitary evolution operator:
\begin{eqnarray}
\label{eq:U}
\hat U(\Delta t) &=& \exp(\hat H \Delta t/{i\hbar}) \cr
&\approx & 1 + (g \hat a_s\hat a_i + h.c.) ~~{\rm when}~~g\equiv \xi \Delta t \ll 1.~~~~
\end{eqnarray}
where we replace the labeling of the fields in Eq.(\ref{eq:H-PA}) by $s,i$, which stand for ``signal, idler" due to historic reason and assume $g\equiv \xi \Delta t$ is a positive number and only keep the first order in the expansion of the exponential. Then with vacuum input, the output state is a two-photon state of the form
\begin{eqnarray}
\label{eq:outstate}
|\Psi\rangle_{PA}= \hat U(t) |vac\rangle
&\approx & |vac\rangle + g \hat a_s^{\dag}\hat a_i^{\dag}|vac\rangle \cr
&=& |vac\rangle + g |1_s,1_i\rangle.
\end{eqnarray}

For the multi-stage interferometer in Fig.\ref{fig:multi-PA}, the output state is then
\begin{eqnarray}
\label{eq:outstate-mPA}
|\Psi\rangle_{mPA} & = & \hat U_N(\Delta t)\hat \Theta(\theta) ... \hat U_2(\Delta t)\hat \Theta(\theta) \hat U_1(\Delta t) |vac\rangle \cr
&\approx &|vac\rangle + \bigg(\sum_{k=1}^{N} g_k e^{i(N-k) \theta} \bigg)|1_s,1_i\rangle,
\end{eqnarray}
where operator $\hat \Theta(\theta)$  adds a total phase of $\theta$ to the signal and idler field together. So, the multi-stage interferometer is equivalent to one PA but with amplitude gain equal to the sum of the amplitude gains of all PAs involved: $g_T=\sum_{k=1}^{N} g_k e^{i(N-k)\theta}$. This is the result of two-photon interference: each PA can generate a pair of photons with amplitude $g_k$ and the final state is a superposition of all the two-photon states.

In the special case when all the PAs have the same gain: $g_k=g$, we have
\begin{eqnarray}
\label{eq:g-mPA}
g_T=g\sum_{k=1}^{N} e^{i(k-1)\theta}  = g e^{i(N-1)\theta/2}H(\theta)
\end{eqnarray}
where $H(\theta)\equiv \frac{\sin N\theta/2}{\sin\theta/2}$ is the multi-path interference factor, which recovers the familiar function of $\cos \theta$ for $N=2$. It first appears in multi-slit interference such as optical grating and has an enhancement factor of $N^2$ for two-photon production rate as compared to single PA. This is the same physics underlying cavity enhanced parametric processes \cite{ou99} and can provide active filtering for spectral mode shaping (see Sect.\ref{sec:VIE} for detail).

The high gain case is not easy to treat because of the general non-commuting nature of the Hamiltonian for different PAs\cite{sipe}. Nevertheless, it still gives rise to the modification of the mode structure at the output similar to the low gain case.

\section{SU(1,1) interferometers of different waves}\label{sec:V}

\subsection{SU(1,1) interferometer with microwaves}\label{sec:VA}

Parametric amplifiers were first realized in radio frequency and microwaves \cite{louisell}. However, thermal and electronic noise is often so large that it overwhelms the quantum noise in detection processes. So, it is hard to study the quantum behavior of the amplifiers in radio frequency and microwave regime. This was changed recently when near quantum limit low noise parametric amplifiers were invented \cite{roc12}. Although thermal and electronic noise is still very high in detection processes, Flurin {\it et al.} \cite{fl12} utilized the low noise parametric amplifier at high gain as a beam splitter to reveal the EPR-type quantum correlation between two entangled microwave fields generated by another low noise parametric amplifier. Similar to the role played for loss-tolerance by the second parametric amplifier in an SU(1,1) interferometer, Flurin {\it et al.}\cite{fl12} used the low noise parametric amplifier to amplify the quantum noise to a level that is much larger than the thermal and electronic background noise in the detection process. In this way, they achieved the measurement of the correlated quantum noise from EPR-entangled microwave fields even in the presence of the enormous thermal and electronic background noise.

\begin{figure}
\begin{center}
\includegraphics[width=3.2 in]{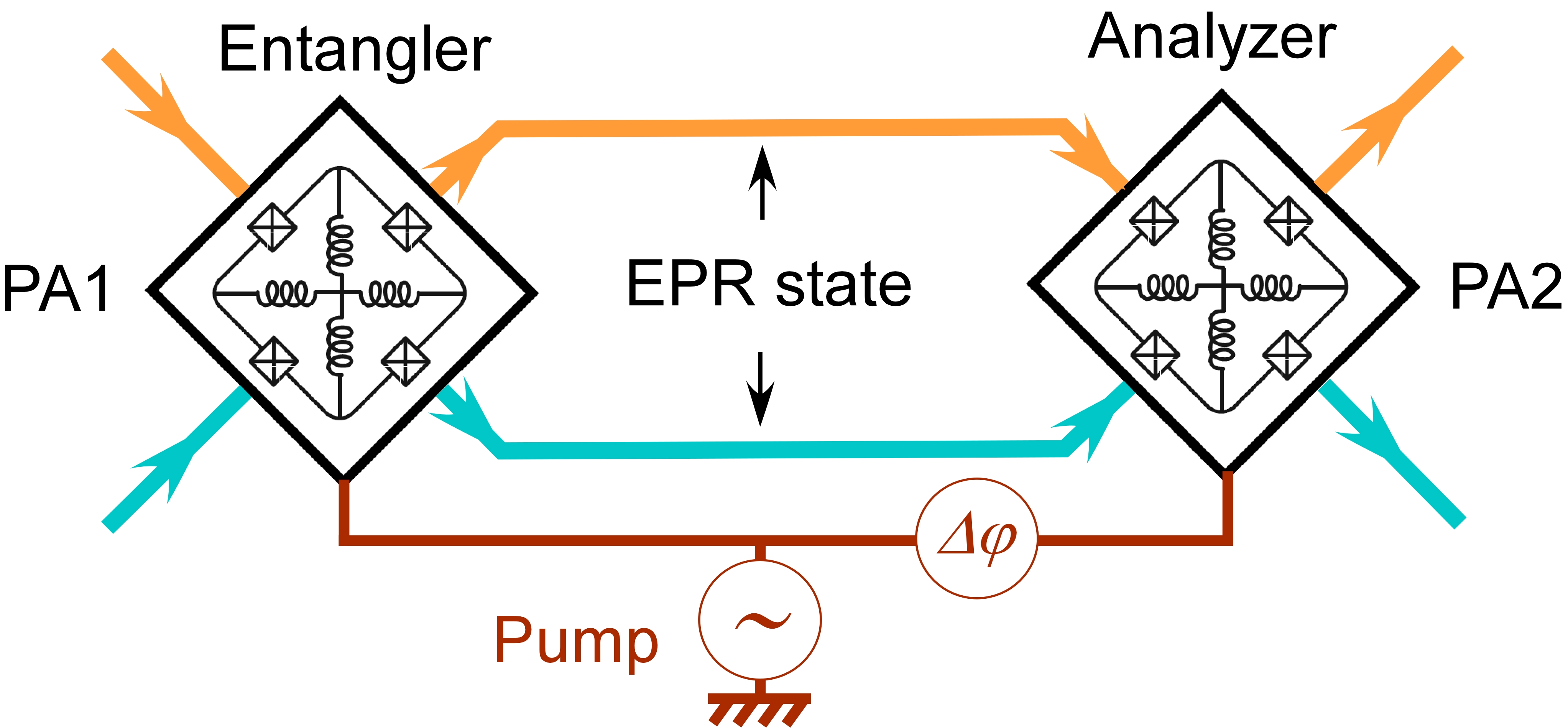}
\end{center}
\caption
{SU(1,1) interferometer for microwaves. Adapted from Ref.\citenum{fl12}. \label{fig9-0} }
\end{figure}

Working on the goal to demonstrate the EPR-type entanglement between microwave fields,  Flurin {\it et al.} \cite{fl12} inadvertently realized an SU(1,1) interferometer in microwave regime. In their arrangement shown in Fig.\ref{fig9-0}, the first amplifier (PA1) is the EPR-entangled source (Entangler) while the second one (PA2) is the one that measures the entanglement (Analyzer). This geometry is exactly in the form of Fig.\ref{SUI} and is an SU(1,1) interferometer but without seeding of a coherent state. Indeed, the measurement result shows an interference pattern that depends on the phase difference $\Delta\varphi$ of the pumps \cite{fl12}. Notice that the required high gain setting for the analyzer amplifier in this case is exactly the setting for achieving the optimum performance of the SU(1,1) interferometer presented in Eq.(\ref{eq:SUI-SNR2}).
\vskip 0.6in

\subsection{Atom-light Hybrid interferometers}\label{sec:VB}

One of the key differences of an SU(1,1) interferometer from a traditional interferometer is the way of wave splitting and superposition for interference: it is through nonlinear mixing of waves. This method can therefore couple different types of waves for interference, which is basically impossible in a traditional interferometer. This leads to hybrid interferometers where the two interfering waves are different types of waves. One such interferometer is the atom-light hybrid interferometer, first realized by Chen {\it et al.} in 2015 \cite{chen15}.

\begin{figure}
\begin{center}
\includegraphics[width=3.2 in]{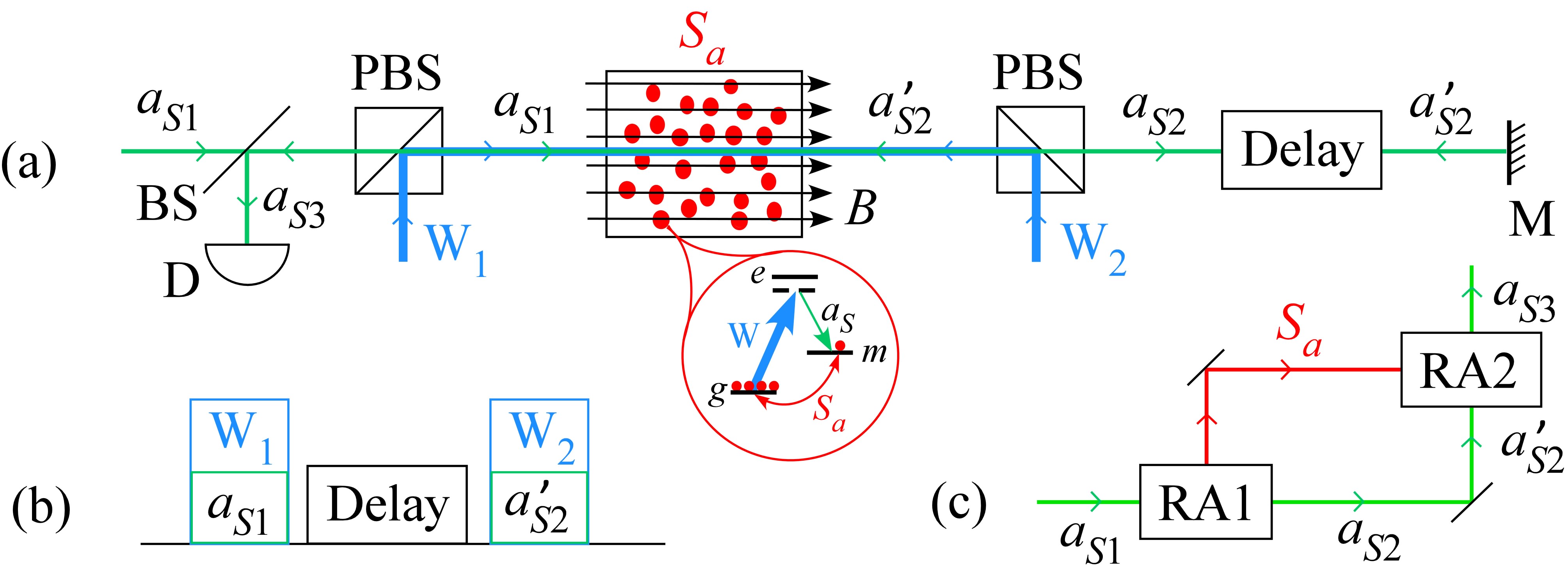}
\end{center}
\caption
{Hybrid atom-light interferometers. (a) Schematic diagram of the interferometer; PBS: polarization beam splitter, BS: beam splitter, M: mirror, D: detector, B: magnetic field for atomic phase change. (b) Time sequence of light pulses. (c) MZ interferometer-equivalent interference paths for atomic spin wave $S_a$ and optical wave $a_S$, RA1,RA2: Raman amplifier. Inset: atomic levels and optical waves. \label{fig-ali} }
\end{figure}

Similar to the all-optical SU(1,1) interferometer in the original realizations \cite{jing11,hud14}, the wave splitting and superposition elements in an atom-light hybrid interferometer are Raman amplifiers which are a special kind of parametric process coupling light waves of strong Raman pump field $A_W$ and Stokes field $\hat a_S$ with an atomic collective excitation wave $\hat S_a$ (also known as pseudo-spin wave) between two lower states ($g,m$) via an excited state ($e$), as shown in the inset of Fig.\ref{fig-ali}. The Raman interaction Hamiltonian\cite{DLCZ,pol} has the same form as the parametric interaction Hamiltonian in Eq.(\ref{eq:H-PA}):
\begin{eqnarray}
\hat H_R=i\hbar \eta A_{W} \hat a_{S}^{\dag}\hat S_a^{\dag}
-i\hbar \eta^* A_{W}^{*}\hat a_S\hat S_a,\label{Hr}
\end{eqnarray}
except that one of the light field, say $\hat a_2$, is replaced by the atomic spin wave $\hat S_a$ and the other field $\hat a_1$ is renamed as the Stokes field $\hat a_{S}$.

In most applications of Raman amplifiers, the atomic states are treated as inaccessible internal states of the amplifier, which are often in the vacuum state (unexcited state) and are not taken into consideration. They are responsible for the spontaneous emission noise of the amplifier. For the action of SU(1,1) interferometer, as we see from the previous section, it  requires the atomic spin wave to participate as one of the interfering fields. Therefore, atomic spin wave is a part of the waves participating in the interference together with the optical Stokes field. So, interference fringe will depend on both the atomic phase and the optical phase, thus forming an atom-light hybrid SU(1,1) interferometer. The schematic diagram is shown in Fig.\ref{fig-ali}(a). The input Stokes field $a_{S1}$, after interacting with atoms pumped by the first writing field $W_1$ (see Fig.\ref{fig-ali}(b) for time sequence), is amplified as $a_{S2}$. In the meantime, an atomic spin wave $S_a$ is also generated in the atomic ensemble. This is the wave splitting process (RA1 in Fig.\ref{fig-ali}(c)). Since the atomic spin wave stays in the atomic ensemble, to combine it with the amplified Stokes, we send back with a mirror (M) the delayed Stokes field $a_{S2}'$ together with the second write field $W_2$ (RA2 in Fig.\ref{fig-ali}(c), see Fig.\ref{fig-ali}(b) for time sequence). The output $a_{S3}$ is detected by D to reveal interference fringe as the optical or atomic phase is scanned. The atomic phase can be changed by external magnetic field via Zeeman effect, as demonstrated by Chen {\it et al.}\cite{chen15}

Atomic phase can also be altered by shining an off-resonant light beam on the atoms via the AC Stark shift \cite{qiu16}.
Thus, an interesting application of the atom-light interferometer is to measure the photon number of the off-resonant light field in the sense of quantum non-demolition (QND) measurement \cite{chen17}. This approach is similar to the QND measurement scheme for microwave photons \cite{haroche}.

\subsection{Atomic SU(1,1) interferometer}\label{sec:VC}

\begin{figure}
\begin{center}
\includegraphics[width=\linewidth]{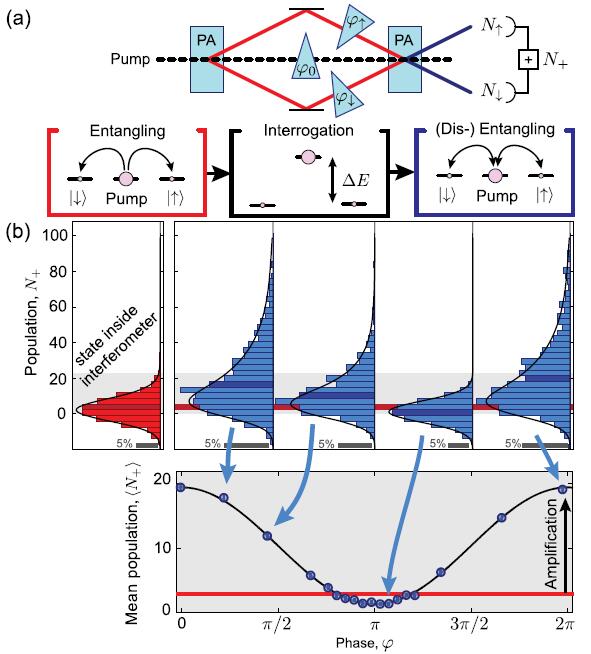}
\end{center}
\caption{Atomic SU(1,1) interferometer. (a) Interferometric scheme with the wave splitting and recombination processes equivalent to parametric amplifiers (PA). (b) The output atomic number distributions as the phase of the atomic waves changes. Reproduced from Ref.~\citenum{BEC}.}
\label{fig11}
\end{figure}

The atomic interferometer discussed in the previous section is a hybrid version involving optical waves in interference.  An all-atom version of the SU(1,1) interferometer was first realized by Linnemann {\it et al.}\cite{BEC} in a spinor Bose-Einstein condensate. A PA+BS variational version (Sect.\ref{sec:IVA}) of the SU(1,1) interferometer with atoms was realized earlier by Gross {\it et al.} \cite{gro10}. The nonlinear interaction responsible for atomic wave splitting and superposition is the spin exchange collision between $^{87}Rb$ atoms of spin $F=2$ manifold and has a Hamiltonian of the form similar to Eq.(\ref{eq:H-PA}) for the parametric process:
\begin{eqnarray}
\hat H_{at}=\hbar \kappa  \hat a_{\uparrow}^{\dag}\hat a_{\downarrow}^{\dag}
+ H.c.,\label{Hat}
\end{eqnarray}
where $\hat a_{\uparrow},\hat a_{\downarrow}$ correspond to the atomic fields in the spin states of $|\uparrow\rangle \equiv |F=2,m_F=1\rangle$ and $|\downarrow\rangle \equiv |F=2,m_F=-1\rangle$, respectively. The effective nonlinear coupling $\kappa \equiv g N_0$ is related to the microscopic nonlinearity $g$, arising from coherent collisional interactions and the number of colliding atoms $N_0$ in the initial state of $|F=2, m_F=0\rangle$, acting as the pump mode. Figure \ref{fig11} shows the schematic of the interferometer (a) and the phase-dependent atomic numbers with their average showing the interference pattern (b). Note that the sum of the two output channels is measured because they are in phase, which is the unique property of SU(1,1) interferometer.  This version of the SU(1,1) interferometer is the unseeded one without coherent state injection since initially there is no atom in either $|\uparrow\rangle$ or $|\downarrow\rangle$ state. Nonetheless, phase measurement sensitivity beyond the SQL was demonstrated.

The atom-light hybrid interferometer discussed in the previous section and the atomic SU(1,1) interferometers discussed here all involve atomic internal states. An atomic interferometer usually refers to interferometers involving the de Broglie matter waves of atoms via their external motional states \cite{prich}. An SUI of this type requires matter wave amplifiers \cite{ketter,deng}, which can be realized by four-wave mixing of matter waves \cite{deng2}.

For the hybrid atom-light interferometer involving the external translational degrees of freedom of atoms, we need to go back to Raman amplification but deal with ultra-cold atoms in a BEC \cite{ino99,koz99,sch04,yosh04} where super-radiance of light is correlated with the atomic motional states in a similar way as in Eq.(\ref{Hr}).

\subsection{Phonon SU(1,1) interferometer}\label{sec:VD}

Parametric amplifiers are the essential ingredients for an SU(1,1) interferometer. Nonlinear interactions are usually involved for them as we have seen before in Raman amplifier and  parametric processes.
Opto-mechanical systems couple light fields with a mechanical oscillator and can realize similar nonlinear interaction for parametric amplification. The opto-mechanical coupling between a mechanical oscillator and a single
optical cavity mode has an interaction Hamiltonian given by \cite{law95}
\begin{eqnarray}
\hat H_{OM}=\hbar \gamma  \hat a^{\dag}\hat a \hat x_m,\label{Hom}
\end{eqnarray}
where $\hat x_m = \hat b+\hat b^{\dag}$, $\hat a$ and $\hat b$ are the annihilation operators for the optical cavity mode and the phonon mode of
the mechanical oscillator, respectively, and $\gamma$ is the opto-mechanical coupling constant. With a strong coherent optical field, we can make a linear approximation: $\hat a = \alpha + \hat a_s$ and the Hamiltonian in Eq.(\ref{Hom}) becomes
 \begin{eqnarray}
\hat H_{OM}\approx \hbar \Gamma  (\hat a_s^{\dag}\hat b + h.c.) + \hbar \Gamma  (\hat a_s^{\dag}\hat b^{\dag}+ h.c.),\label{Hom2}
\end{eqnarray}
where $\Gamma=\gamma\alpha$ is the effective opto-mechanical coupling rate. The first term in Eq.(\ref{Hom2}) has the form of the well-known beam-splitter Hamiltonian whereas the second term is similar to a parametric amplification process given in Eq.(\ref{eq:H-PA}). The derivation above is oversimplified without considering multi-mode nature of the optical field. With a multi-mode model, the interaction can be viewed as a Raman process so that the second term in Eq.(\ref{Hom2}) corresponds to the Stokes scattering while the first term to the anti-Stokes scattering. Whichever term dominates the interaction depends on the cavity resonance to Stokes or anti-Stokes component of the optical field.
In analogy with a Ramsey interferometer, Qu {\it et al.}\cite{qu14} utilized the beamsplitter-like Hamiltonian in the first term of Eq.(\ref{Hom2}) to realize an opto-mechanical Ramsey interferometer. Of course, had they use the second term of Eq.(\ref{Hom2}), it would become a hybrid photon-phonon SU(1,1) interferometer in the same spirit of the atom-light hybrid interferometer discussed in Sect.\ref{sec:VB}.

\begin{figure*}
\begin{center}
\includegraphics[width=4.7 in]{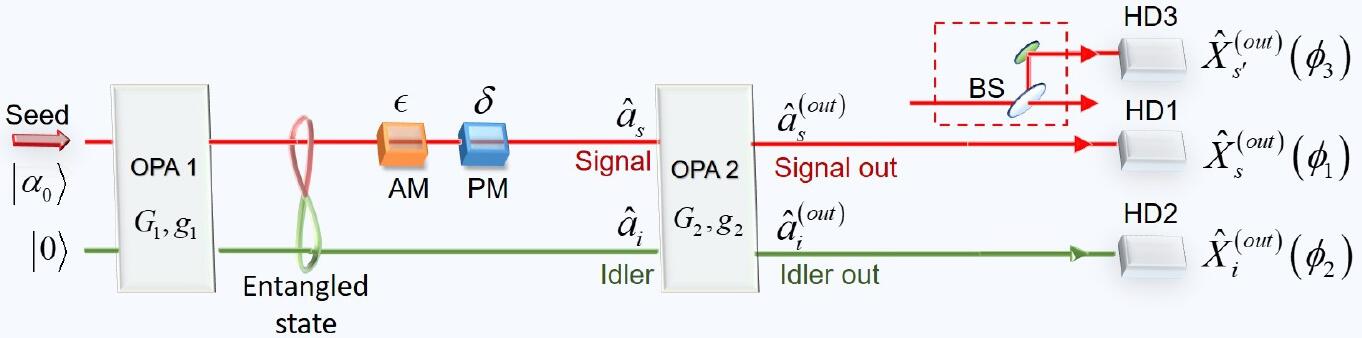}
\end{center}
\caption{Simultaneous phase ($\delta$) and amplitude ($\epsilon$) measurement by an SU(1,1) interferometer. HD: homodyne detection, BS: beam splitter. Reproduced from Ref.\citenum{liu19}.}
\label{fig-multi}
\end{figure*}

For an all-phonon S(1,1) interferometer, strong nonlinear interaction between mechanical oscillators is required. This was realized by Patil {\it et al.}\cite{pat15}, who demonstrated parametric amplification of phonons and thermo-mechanical noise squeezing. Equipped with phonon parametric amplifiers, Cheung {\it et al.} \cite{OM} realized a PA+BS version of all-phonon SU(1,1) interferometer (Sect.\ref{sec:IVA}) where the second parametric amplifier is replaced by a beam splitter.

\section{Applications of SU(1,1) interferometers}\label{sec:VI}

The primary application of SU(1,1) interferometers is in phase measurement. It was shown in Sect.\ref{sec:IIID} that the phase measurement sensitivity can beat the standard quantum limit. On the other hand, as we have found in Sect.\ref{sec:IIIC1}, the noise reduction in SUI is due to quantum destructive interference, which reduces all noise of the whole output fields of the interferometer. So, the sensitivity enhancement effect is not limited to phase measurement and can also be applied to the measurement of other quadrature-phase amplitudes such as amplitude measurement, as we will show next.

\subsection{Multi-parameter measurement}\label{sec:VIA}

As is well-known, phase and amplitude are conjugate variables so that Heisenberg uncertainty principle prevents their simultaneous measurement with precision beyond what the uncertainty principle allows. However, this limitation is on one object and entanglement between two objects can break this limitation, which is what leads to the famous EPR paradox in an apparent violation of Heisenberg uncertainty relation \cite{reid,ou92,reid09}. But Braunstein and Kimble \cite{brau} made use of this seemingly contradicting behavior of the entangled source to achieve simultaneous measurement of phase and amplitude with measurement precision beating the limit set by Heisenberg uncertainty relation. The experimental demonstration of this phenomenon was first performed by Li {\it et al.} \cite{li02} following a proposal by Zhang and Peng \cite{zhang}, which is a variation of the quantum dense coding scheme of Braunstein and Kimble \cite{brau}. More recently, Steinlechner {\it et al.} \cite{snb13} applied the same technique to a prototype interferometer for gravitational wave detection with simultaneous measurement of two non-orthogonal quantities.

A look at the quantum dense coding scheme by  Braunstein and Kimble reveals that it is  just the PA + BS scheme of the SU(1,1) interferometer that we discussed in Sect.\ref{sec:IVA}. Since the role of the BS is the same as the second parametric amplifier in SU(1,1) interferometer to superpose the two entangled fields, the original SU(1,1) interferometer with two parametric amplifiers should be able to accomplish the same task as the quantum dense coding scheme. Indeed, it was theoretically shown \cite{JML18} that while homodyne detection of $\hat Y_1$ at the signal output port  of PA2 (port 1 in Fig.\ref{SUI}) gives rise to the measurement of the phase modulation $\delta$ with an SNR of
\begin{eqnarray}
\label{SNR-ph}
SNR_{Ph}^{(1)}
&=& \frac{4g_2^2I_{ps}\delta^2}{(G_1^2+g_1^2)(G_2^2+g_2^2) -4G_1G_2g_1g_2} \cr
&\rightarrow & 2(G_1+g_1)^2 I_{ps}\delta^2 ~~~~{\rm for}~~~~g_2\rightarrow \infty,
\end{eqnarray}
as presented in Eq.(\ref{eq:SUI-SNR-1}), homodyne detection of $\hat X_2$ at the idler output port of PA2 (port 2 in Fig.\ref{SUI}) leads to the measurement of the amplitude modulation with an SNR of
\begin{eqnarray}
\label{SNR-am}
SNR_{Am}^{(2)}
&=& \frac{4G_2^2I_{ps}\epsilon^2}{(G_1^2+g_1^2)(G_2^2+g_2^2) -4G_1G_2g_1g_2}\cr
&\rightarrow & 2(G_1+g_1)^2 I_{ps}\epsilon^2~~~~{\rm for}~~~~g_2\rightarrow \infty,
\end{eqnarray}
where $\epsilon$ is the amplitude modulation signal. Similar to the phase measurement sensitivity given in Eq.(\ref{SNR-ph}), the sensitivity of the amplitude measurement presented in Eq.(\ref{SNR-am}) also beats the standard quantum limit.  Figure \ref{fig-multi} shows the schematic of the SU(1,1) interferometer for the simultaneous measurement of both a phase shift and an amplitude change on the signal field.
Note that the phase measurement (HD1 for $\hat X_s(\phi_1)$ with $\phi_1=\pi/2$) and amplitude measurement (HD2 for $\hat X_i(\phi_2)$ with $\phi_2=0$) are performed at different ports (signal and idler output ports) that are independent of each other. Therefore, we can make the simultaneous measurement of phase and amplitude with their sensitivities simultaneously beating the standard quantum limit.

The above application of SU(1,1) interferometer to the joint measurement of phase and amplitude was experimentally demonstrated by Liu {\it et al.}\cite{liu19} Furthermore, the measurement scheme is extended to simultaneous measurement of multiple non-commuting observables which are not necessarily orthogonal (HD1 for $\hat X_s(\phi_1)$, HD2 for $\hat X_s(\phi_2)$, HD3 for $\hat X_{s'}(\phi_3)$ in Fig.\ref{fig-multi} with arbitrary $\phi_1, \phi_2, \phi_3$). Since outputs are amplified, we can further split the signal output without introducing vacuum noise for the simultaneous measurement of another modulation non-orthogonal to phase and amplitude by HD3 of $\hat X_{s'}(\phi_3)$.

The advantages of the SU(1,1) interferometer over the quantum dense coding scheme \cite{brau} are: (1) more than two non-commuting quantities can be simultaneously measured and (2) it is tolerant to propagation and detection losses.

\subsection{Quantum resource sharing}\label{sec:VIB}

When using SU(1,1) interferometers for the simultaneous measurement of phase and amplitude, there exists an interesting relation between the optimum sensitivities of the two measurements.  Expressed in terms of the signal-to-noise ratios, the relation is written as \cite{JML18,ass20}
\begin{eqnarray}
SNR_{Ph} + SNR_{Am} = SNR_{op},\label{QRS}
\end{eqnarray}
where $SNR_{op}$ is the optimized SNR of the corresponding  measurement when the resource is all devoted to that measurement so that it is impossible to make the other measurement. This can be seen from Eqs.(\ref{SNR-ph},\ref{SNR-am}) where, if we set the modulation signals equal: $\delta=\epsilon$ and add the two SNRs, we have
\begin{eqnarray}
SNR_{Ph}^{(1)} + SNR_{Am}^{(2)} = 4(G_1+g_1)^2I_{ps}\delta^2 = SNR_{op},\label{QRS2}
\end{eqnarray}
where $SNR_{op}$ is given by Eq.(\ref{eq:SNR-dual-JM}) for the optimum phase measurement sensitivity obtained in the dual-beam scheme. In fact, the less-than-optimized results in Eq.(\ref{eq:SNR-dual1}) for finite $g_1$ in the dual beam scheme can be attributed to the relation in Eq.(\ref{QRS2}) of quantum resource sharing. When $Y_1$ is measured for phase modulation signal, the other port can still be used for amplitude measurement but a straightforward calculation gives an SNR of
\begin{eqnarray}
\label{eq:SNR-dual-am}
SNR_{Am}^{(2)} = 2 I_{ps}\epsilon^2/(G_1^2+g_1^2),
\end{eqnarray}
where we set $g_2\rightarrow \infty$ for optimum value and $\epsilon$ is the amplitude modulation signal. This, when set to have $\epsilon =\delta$, together with the SNR for phase modulation in Eq.(\ref{eq:SNR-dual1}) leads to Eq.(\ref{QRS2}) for quantum resource sharing even at finite $g_1$. Furthermore, although the phase measurement result of the joint measurement between the two ports in Eq.(\ref{eq:SNR-dual-JM}) gives $SNR_{op}$, since both ports are used for phase measurement, which leaves no room for amplitude measurement, namely  $SNR_{Am} =0$, this again satisfies Eq.(\ref{QRS2}) for quantum resource sharing.
Note from Eq.(\ref{eq:SNR-dual-am}) that when $g_1\rightarrow \infty$, $SNR_{Am}^{(2)}\rightarrow 0$, indicating that the dual-beam scheme discussed in Sect.\ref{sec:IVC} is not suitable for amplitude modulation measurement. The reason for this can be traced to the intensity correlation between the two entangled fields from the first parametric amplifier, or the so-called twin beam effect \cite{kumar}: noise in the intensity difference is reduced due intensity correlation whereas the amplitude modulation signal encoded in the two fields are also canceled, leading to no amplitude modulation signal in the intensity difference.

\subsection{Quantum information tapping}\label{sec:VIC}

It is well-known \cite{sha80} that when quantum information is split with a beam splitter, vacuum noise comes in from the unused port, leading to degradation of SNRs of the split signals as compared to the input. Shapiro suggested using squeezed states to combat the vacuum noise and preserve the SNRs of the split signals \cite{sha80}. This is the so-called quantum information tapping. Such a scheme was implemented by Bruckmeier {\it et al.}\cite{bru97}. 

The SU(1,1) interferometer discussed here can be used for quantum information tapping. Consider the two outputs of PA2. We have from Eqs.(\ref{eq:SUI-SNR2}) for $g_2\rightarrow \infty$
\begin{equation}
\label{SNR-QIT}
SNR_{SUI}^{(1)} = SNR_{SUI}^{(2)} = 2(G_1+g_1)^2I_{ps}\delta^2.
\end{equation}
So, the two outputs are identical copies of each other, which can be thought of as the two split signals for the modulated phase signal encoded to the input field before PA2. The input SNR is obtained from the direct measurement and is given as $SNR_{in} = 2(G_1+g_1)^2\delta^2I_{ps}$.\cite{JML18} Hence, we have the transfer coefficients, which are defined as ${\cal T}^{(1,2)} \equiv SNR_{SUI}^{(1,2)}/SNR_{in}$, satisfying the relation for quantum information tapping:
\begin{equation}
\label{T-QIT}
{\cal T}^{(1)} + {\cal T}^{(2)} = 2.
\end{equation}
Notice that classical tapping limit is ${\cal T}^{(1)} + {\cal T}^{(2)} \le 1$.\cite{sha80} The experimental implementation of the quantum information tapping scheme from an SU(1,1) interferometer was realized by Guo {\it et al.}\cite{guo16}, which is basically the amplifier version \cite{lev93} of quantum information tapping but with quantum entangled fields as the input in order to achieve noiseless quantum amplification \cite{ou93,kong13b}. A variation of this scheme is the much improved dual-beam encoding scheme in an SU(1,1) interferometer \cite{liu18}. It was demonstrated for the first time by the quantum information tapping technique that a quantum enhanced signal can be split into two while still maintaining the quantum enhancement property. The SUI scheme of quantum information tapping was extended by Liu {\it et al.} \cite{liu16} to a three-way quantum information tapping scheme for quantum information cascading.

\subsection{Measurement of Entanglement in Continuous Variables}\label{sec:VID}

Verification of quantum entanglement between two light sources is a basic experimental technique in quantum information. For continuous variables, it is usually done via homodyne detection technique by directly measuring the quadrature-phase amplitude correlations of the two fields: $\langle\Delta^2\hat X_-\rangle$ and $\langle\Delta^2\hat Y_+\rangle$ with $\hat X_-\equiv {\hat X_1} - {\hat X_2}$ and $\hat Y_+\equiv {\hat Y_1} + {\hat Y_2}$. Quantum entanglement satisfies the inseparability criterion: $I \equiv\frac{1}{4}(\langle \Delta^2\hat X_-\rangle + \langle \Delta^2\hat Y_+\rangle)<1$.\cite{duan00} However, this traditional homodyne method is prone to loss, which severely limits the application of entanglement. On the other hand, as we have shown in Sect.\ref{sec:IIIA}, parametric amplifiers (PA) can act as non-conventional beam splitters for mixing of two fields, which is exactly performed  when quantities $\hat X_-\equiv {\hat X_1} - {\hat X_2}$ and $\hat Y_+\equiv {\hat Y_1} + {\hat Y_2}$ are measured, forming an SU(1,1)-type interferometer. We analyze this scheme next.

\begin{figure}
\begin{center}
\includegraphics[width=3.2in]{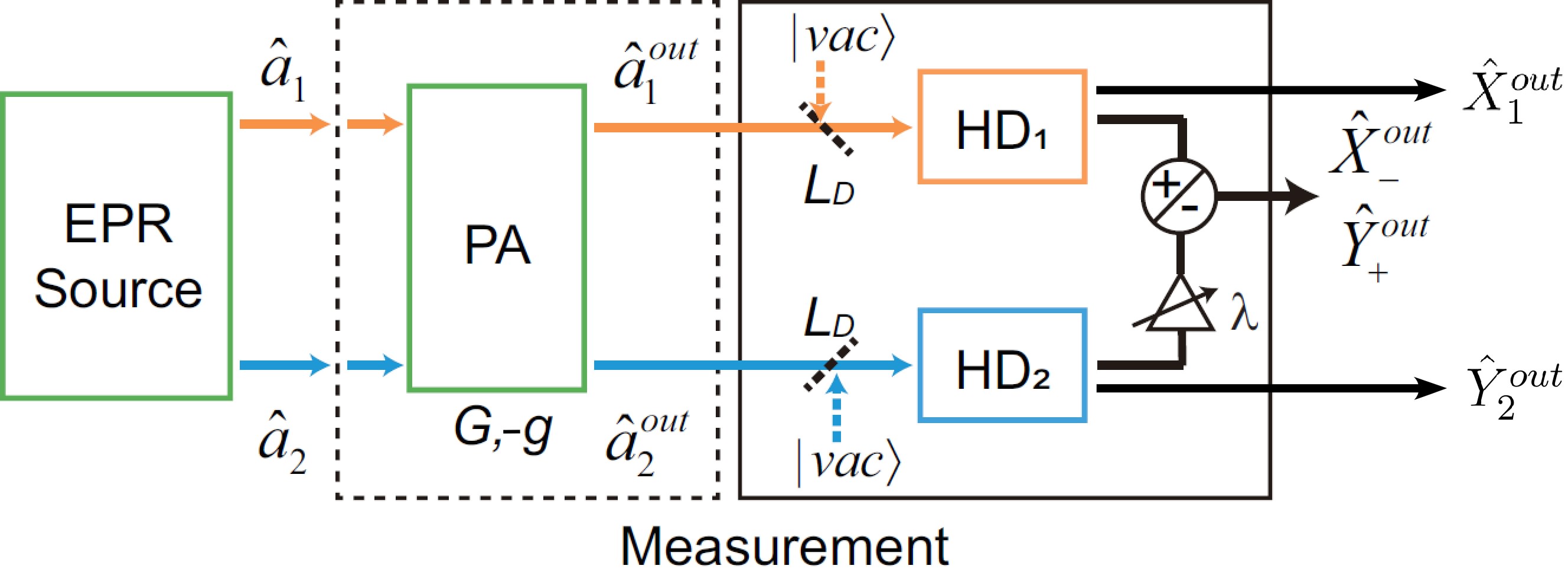}
\end{center}
\caption{Entanglement measurement with the help of a parametric amplifier (PA). HD: homodyne detection. $L_D$: detection loss. Reproduced from Ref.~\citenum{li19}.}
\label{fig12}
\end{figure}

As shown in Fig.\ref{fig12}, the two fields $\hat a_1$ and $\hat a_2$, whose entanglement property needs to be characterized, enter the input ports of a parametric amplifier (PA) of amplitude gain parameters $G,g$. We perform homodyne detections (HD1,HD2) at the two outputs: one for $X$-quadrature ($\hat X_1^{(o)}$), the other for $Y$-quadrature ($\hat Y_2^{(o)}$), similar to the scheme of joint measurement of phase and amplitude. According to Eq.(\ref{eq:XYin-out}), we obtain the input and output relations for the $X, Y$-quadratures as
\begin{equation}
\label{eq:XYin-out-g}
\hat X_{1,2}^{(o)} = G \hat X_{1,2} - g \hat X_{2,1},~~\hat Y_{1,2}^{(o)} = G \hat Y_{1,2} + g \hat Y_{2,1}
\end{equation}
where we dropped the input label $(in)$ for clarity and add a $\pi$ phase to $g$ so that it changes sign. Then the results of measurement are
\begin{eqnarray}\label{EPR-PA}
 \langle \Delta^2 \hat X_1^{(o)} \rangle &=& \langle {\Delta^2 (G \hat X_1 - g \hat X_2)} \rangle \cr &=& G^2  \langle {\Delta^2 ( \hat X_1 - k \hat X_2)} \rangle = G^2 \langle {\Delta ^2}\hat X_-^{(k)} \rangle, \cr
  \langle \Delta^2 \hat Y_1^{(o)} \rangle &=& \langle {\Delta^2 (G \hat Y_1 + g \hat Y_2)} \rangle \cr &=& G^2 \langle {\Delta^2 (\hat Y_1 + k\hat Y_2)} \rangle  = G^2 \langle {{\Delta ^2}\hat Y_+^{(k)}} \rangle,
\end{eqnarray}
where $k\equiv g/G \rightarrow 1$ for large $G$ and $\hat X_-^{(k)}\equiv \hat X_1 - k \hat X_2, \hat Y_+^{(k)}\equiv \hat Y_1 + k \hat Y_2$. If we block the two inputs and make measurement for vacuum input, we can obtain the result of uncorrelated vacuum for comparison. Take the ratio for the two measurements, we have
\begin{eqnarray}\label{I-PA}
  I_{amp}^{(1,2)} & \equiv  &\frac{\langle \Delta^2 \hat X_{1,2}^{(o)} \rangle + \langle \Delta^2 \hat Y_{1,2}^{(o)} \rangle }{\langle \Delta^2 \hat X_{1,2}^{(o)} \rangle_{v}+\langle \Delta^2 \hat Y_{1,2}^{(o)} \rangle_{v}} \cr &= &\frac{\langle \Delta^2\hat X_-^{(k)}\rangle + \langle \Delta^2\hat Y_+^{(k)}\rangle }{2(1+k^2)} \rightarrow I ~~{\rm for}~~k\rightarrow 1.
  \end{eqnarray}
Therefore, we can make direct measurement of the inseparability quantity $I$ with a high gain parametric amplifier ($k\rightarrow 1$). The advantage is its tolerance to any loss at detection ($L_D$), as in all applications of SU(1,1) interferometer. The other advantage is the simultaneous measurement of $\hat X_-, \hat Y_+$ with no parameters adjustment. The disadvantage is the need to have a relatively high gain.

Furthermore, if we make joint measurement for the quantities $\hat X_-^{(o)}\equiv \hat X_1^{(o)} - \lambda \hat X_2^{(o)}$ and $\hat Y_+^{(o)}\equiv \hat Y_1^{(o)} + \lambda \hat Y_2^{(o)}$ at the two outputs, one at a time, it can be shown \cite{jml19pra} that with a proper adjustment of the electronic coefficient $\lambda = (kG-g)/(G-kg)$, we can always obtain
\begin{equation}\label{I-amp-JM}
\begin{split}
  I_{amp}^{JM} & = \frac{\langle \Delta^2 \hat X_-^{(o)} \rangle + \langle \Delta^2 \hat Y_+^{(o)} \rangle } {\langle \Delta^2\hat X_-^{(o)}\rangle_{v}+\langle \Delta^2\hat Y_+^{(o)}\rangle_{v}} = \frac{\langle \Delta^2\hat X_-^{(k)}\rangle + \langle \Delta^2\hat Y_+^{(k)}\rangle }{2(1+k^2)} = I^{(k)}
  \end{split}
\end{equation}
for any gain parameters $G,g$. Especially, we have $I_{amp}^{JM} = I$ with $\lambda=1$.  When $G=1, g=0$, this is exactly the method of direct homodyne measurement, but with the help of a parametric amplifier, the scheme is immune to losses.

The discussion above is for single-mode case. For multi-mode case, parametric amplifiers has the ability to select out the dominating mode \cite{jml19pra}. This is because different modes have different parametric gains. In the high gain limit, the mode with largest gain will dominate. Thus application of parametric amplifiers to entanglement measurement can also filter out unwanted higher order modes and concentrate on the dominating mode. By using mode engineering technique on parametric amplifiers discussed in next section, we can select the mode of our interest.

The scheme discussed above for entanglement measurement was implemented experimentally by Li {\it et al.} \cite{li19} with fiber optical parametric amplifiers, demonstrating the loss-tolerant and mode selecting properties for the high gain case.

\subsection{Mode engineering of quantum states with SU(1,1) interferometers}\label{sec:VIE}

Another interesting application of SU(1,1) interferometers is the modification of the mode structure of the quantum state in the output fields to achieve quantum state engineering. The mode structure of quantum states has recently attracted a lot of attentions because it increases the degrees of freedom for quantum fields and is especially appealing to quantum information science because of its ability to achieve multi-dimensional quantum entanglement (see a comprehensive review by Fabre and Treps \cite{fab19}). Modes of photons play essential roles in quantum interference because they define the identity of photons and often lead to distinguishability \cite{ou17}. It is crucial to have mode match between interfering fields in order to achieve high visibility. Our discussion so far on SU(1,1) interferometers has assumed perfect mode match. Multi-mode behavior of SUI is also quite different from linear interferometers. In fact, a recent work demonstrated the mode cleaning ability of SUI \cite{li19,jml19pra}.  In the following, we will discuss the ability of SUI for tailoring the mode structure of quantum fields to our need.

The modification of the mode structure is achieved by engineering the phase change in between the two PAs. This idea was put into action \cite{isk15-1,isk15-2} soon after the first experimental realization of the SU(1,1) interferometers \cite{jing11,hud14}. It can be used to modify both spatial \cite{isk15-1,fra19-1,fra19-2} and temporal/spectral \cite{isk15-2,lem16,sha18,su19,horo20,huo20} profiles of the output fields. More recently, a multi-stage SUI (Sect.\ref{sec:IVD}) was implemented for precise and versatile engineering of the spectral mode function of the output quantum states \cite{jml20}. As an example, we present here the interferometric scheme to achieve flexible and precise spectral mode engineering for the two-photon state produced from spontaneous parametric emission processes (SPE).

When pumped by an ultra short pulse ($\sim$ 100 fs), SPE processes generate a broadband two-photon state of the form similar to Eq.(\ref{eq:outstate}) but with multi-frequency mode description:
\begin{eqnarray}\label{2ph-w}
 |\Psi_2 \rangle = |vac\rangle + g \int d\Omega_s d\Omega_i F(\Omega_s,\Omega_i)\hat a_s^{\dag}(\Omega_s)\hat a_i^{\dag}(\Omega_i)|vac\rangle,~~~~~~~~
\end{eqnarray}
where  $F(\omega_s,\omega_i)$ is the normalized joint spectral function (JSF) describing the joint spectral properties of the signal (subscript $s$) and idler (subscript $i$) photons and has the form of
\begin{eqnarray}\label{JSF}
F(\Omega_s,\Omega_i)= {\cal N} e^{-(\Omega_s+\Omega_i)^2/2\sigma_p^2} {\rm sinc} (\Delta k L/2)
\end{eqnarray}
with $\Delta k$ as the wave vector (phase) mismatch among all the waves in a nonlinear medium of length $L$. $g (\ll 1)$ is similar to the same quantity in Eq.(\ref{eq:outstate}) and is proportional to $L$ and nonlinear coefficient and is related to the peak amplitude of the pump field. ${\cal N}$ is the normalization constant.  Note that Eq.(\ref{JSF}) is written in terms of the frequency offsets $\Omega_s, \Omega_i$ from the central frequencies $\omega_{s0},\omega_{i0}$ of the generated fields which are determined by phase matching condition $\Delta k =0$ and the center frequency of the pump field.

\begin{figure}
\begin{center}
\includegraphics[width=3.1in]{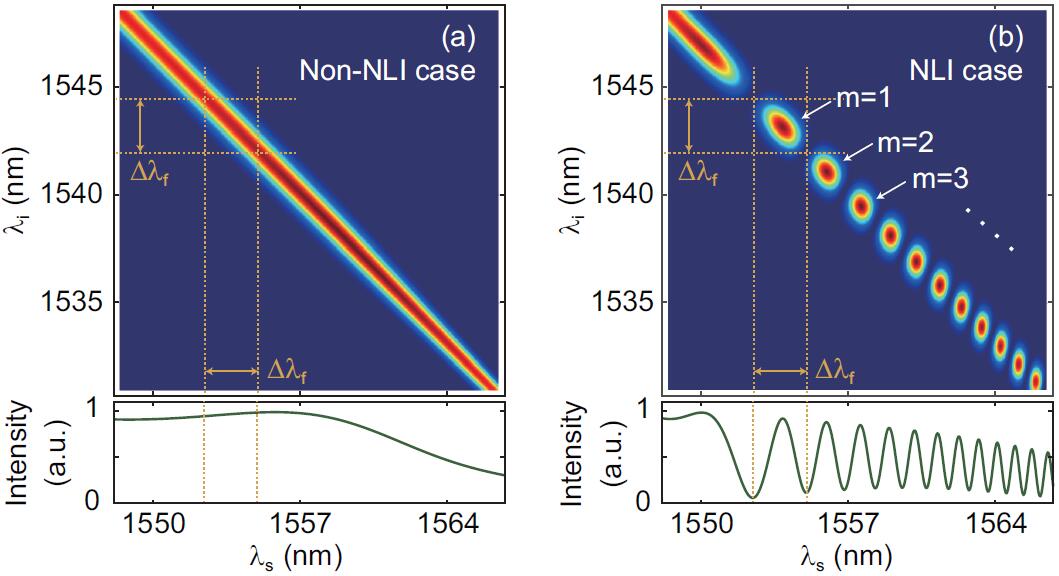}
\end{center}
\caption{Contour plot of the joint spectral function $F(\Omega_s,\Omega_i)$ (JSF). (a) A single parametric process; (b) An SU(1,1) interferometer with spectrally dependent phase for the modification of JSF. Marginal intensity distribution $I(\omega_s)$ below the horizontal axes. Reproduced from Ref.~\citenum{su19}.}
\label{Fig-JSF}
\end{figure}

If the frequencies of the signal and idler photons are close to each other, that is,  $|\omega_{s0}-\omega_{i0}|\ll \omega_{s0},\omega_{i0}$, then the sinc-function in $F(\Omega_s,\Omega_i)$ has a broad bandwidth much wider than the pump bandwidth $\sigma_p$. In this case, $F(\Omega_s,\Omega_i)$ is mainly determined by the exponential function and forms a strip along $-~45^{\circ}$, as shown in Fig.\ref{Fig-JSF}(a). The strip orientation of $-~45^{\circ}$ reflects the frequency anti-correlation between the signal and idler photons due to energy conservation of photons. This shape of JSF gives rise to a two-photon state with multiple temporal modes and is not desirable for quantum interference in quantum information applications. Single-mode two-photon states are preferable, corresponding a factorable JSF with a shape such as a round circle.

The JSF of two-photon state can be modified by an SU(1,1) interferometer with a spectrally dependent phase $\theta$ by using a linear dispersive medium sandwiched in between the two PAs, as shown in Fig.\ref{fig:multi-PA} with $N=2$. Then the interference term $H(\theta)= 2\cos \theta $ in Eq.(\ref{eq:g-mPA}) will modify the single PA term $g$, which becomes the JSF $F(\Omega_s,\Omega_i)$ in Eq.(\ref{JSF}) for the broadband case. For the case of near degenerate frequencies of $|\omega_{s0}-\omega_{i0}|\ll \omega_{s0},\omega_{i0} $, the first-order dispersion disappears and second order dispersion leads to $\theta = \beta (\Omega_s-\Omega_i)^2L_{DM}$ with $\beta$ proportional to second order dispersion coefficient and $L_{DM}$ as the length of the linear dispersive medium. Figure \ref{Fig-JSF}(b) shows the modified JSF together with the marginal intensity $I(\omega_s)$ of the signal field, showing the interference fringe. The island structure of the modified JSF is a result of two-photon interference and forms a multi-dimensional two-photon state with entanglement between different islands \cite{mon}. Filters can be used to select the roundest island for a nearly factorable JSF (dashed yellow lines). The shape of the islands can be adjusted depending on the pump bandwidth $\sigma_p$ and the length $L_{DM}$ of the dispersive medium.

\begin{figure}
\begin{center}
\includegraphics[width=3.1in]{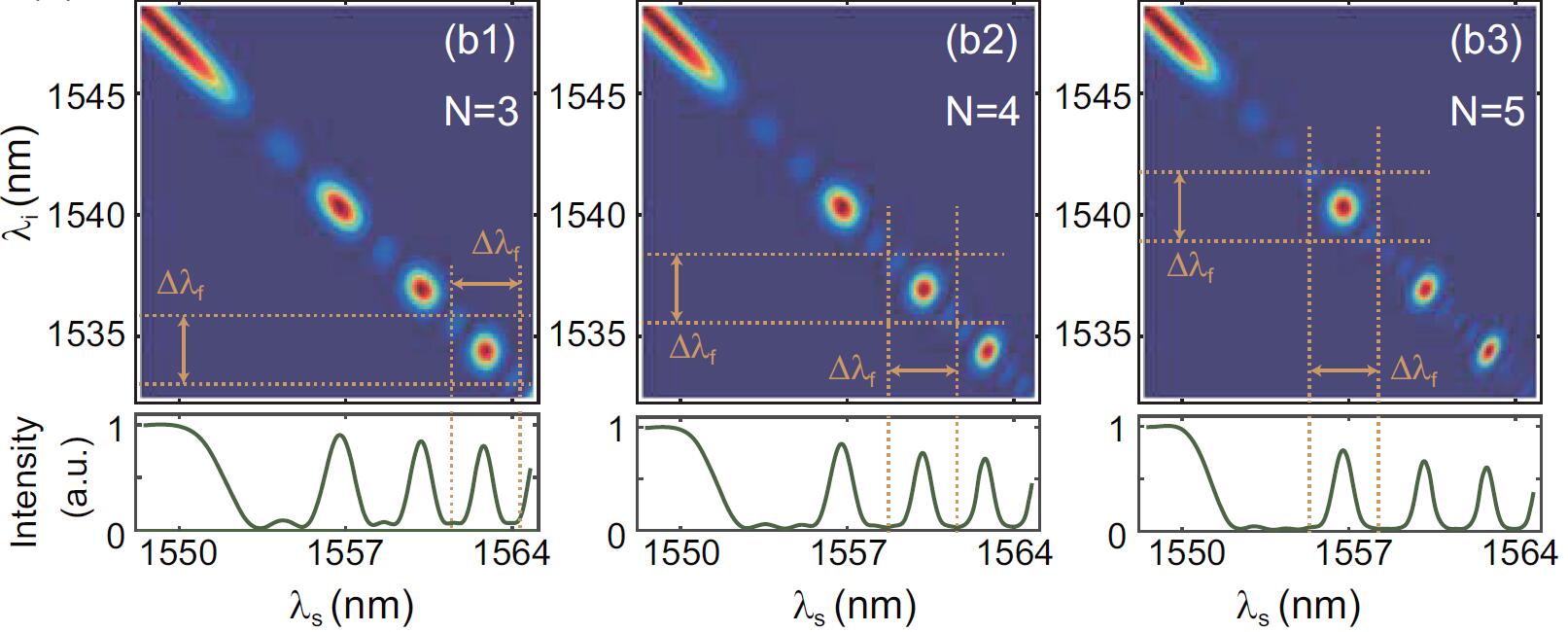}
\end{center}
\caption{Contour plot of the joint spectral function $F(\Omega_s,\Omega_i)$ (JSF) for the output state from a multi-stage SU(1,1) interferometer. $N=$ (b1)3, (b2) 4, (b3) 5. Reproduced from Ref.~\citenum{su19}.}
\label{Fig-JSF2}
\end{figure}

The cleanliness and thus better quality of the selected island depends on the visibility of interference. This can be improved with a multi-stage design, presented in Sect.\ref{sec:IVD}, where the interference term $H(\theta)=\frac{\sin N\theta/2}{\sin \theta/2}$ will make the islands well separated with increasing stage number $N$, as shown in Fig.\ref{Fig-JSF2}. The improved visibilities in the interference pattern shown in the marginal intensity should give a cleaner JSF with better quality as $N$ increases. However, the well-known mini-peaks of $H(\theta)$ function between the main islands are troublesome. Further shaping of $H(\theta)$ function can be done with a design of uneven gain $g_k$ or gain medium length $L_k (g_k \propto L_k)$ distribution among $N$ PAs. It was shown that a binomial distribution $L_k = L_1 (N-1)!/(k-1)!(N-k)!$ ($k=1,2,...N$) will eliminate the mini peaks, leading to improved JSF\cite{su20}.

With SU(1,1) interferometers, there are many of degrees of freedom for adjustment and fine tuning in the modification of the JSF of the two-photon state to achieve what we want. Although the discussion is for low gain case ($g\ll 1$), it was shown experimentally that the interferometric technique works equally well in the high gain regime for the precise engineering of the mode structure of entangled fields in continuous variables \cite{lem16,sha18,huo20,su20}. But all the theoretical treatment \cite{sha18,su20} in this case does not consider the non-commuting property of the Hamiltonian between different PAs \cite{sipe}.

\section{SUMMARY AND FUTURE PROSPECTS}\label{sec:VII}

SU(1,1) interferometers are a new type of interferometers that employ nonlinear interactions such as parametric processes to split and mix beams for interference. They possess some unique properties, making them advantageous over the traditional beam splitter-based interferometers. These properties include higher sensitivity, detection loss tolerance, and mixing of different types of waves. A key feature of the interferometer is the quantum correlation between the two interfering arms of the interferometer, which is responsible for the enhancement of phase measurement sensitivity. The involvement of parametric amplifier in the superposition of the interfering waves leads to the loss tolerance property that has some practical implications in quantum metrology. The nonlinear mixing of different types of waves for interference opens up doors for potentially much wider application of this new type of interferometer than the traditional interferometers. Despite these demonstrated advantages, we still have many challenges, both fundamental and technological, in the further development of the technique of SU(1,1) interferometers.

Among these advantageous properties, the ability to mix waves of different types in SU(1,1) interferometers will make them more promising than others for sensing applications in wide areas. It will be especially attractive to those waves that lack efficient way of detection such as  THz and far infra-red waves. SU(1,1) interferometers allow sensing of phase change in these waves but make detection at other waves for which detection efficiency is high, so long as there is a coupling between these waves for nonlinear mixing. This should also widen our capability to construct sensors for measuring a variety of physical quantities through these waves. For example, coupling atomic de Broglie wave through translational degrees of freedom to light by super-radiance \cite{sch04,yosh04,haine} will allow us to sense gravitational field.  An all-matter wave SU(1,1) interferometer can also be used to measure gravity and will require matter wave mixing \cite{deng2} for its realization. To make these applications possible, we need to look for nonlinear mixing between waves of our interest. Of course, these interactions may not be in the form of parametric interaction given in Eq.(\ref{eq:H-PA}) and the interferometers constructed with them will not be SU(1,1)-type as we discussed in this paper. They will have totally different properties yet to be explored. An example is the engineered multi-particle interaction for phonons in trapped ion systems \cite{wine}.

The experimental realizations discussed in the paper are mostly proof-of-principle experiments and they are operated under relatively small phase sensing photon number ($I_{ps}$).  For their wide applications in sensing, we still need to see how they can be adapted to practical situations and different environments.
For example, for surpassing the performance of traditional interferometers in actual sensing applications, we need to increase the absolute sensitivity. This is achieved by increasing the phase sensing photon number $I_{ps}$. However, this usually leads to saturation of the parametric amplifiers and other unwanted nonlinear effects such as self-phase modulation in optical fibers. Perhaps the solution to this problem lies in the selection of operating points at relatively low overall gain of the interferometer with double injection as suggested in Ref.\citenum{pl} and realized in Ref.\citenum{li02}.
Different applications require different variations of SU(1,1) interferometers. For example, how can SU(1,1) interferometers be adapted to measure rotation like Sagnac interferometers do? This is not obvious since the SU(1,1) interferometers depend on phase sum instead of phase difference.

To take the quantum advantages for realization of quantum sensing, we need to have an effective way to
control the internal losses of the interferometers. Although SU(1,1) interferometers are relatively immune to external losses such detection inefficiency, what limits the enhancement of sensitivity is the internal losses experienced by the fields in between the two PAs or the losses suffered by the PAs \cite{ou12,mar12}. These losses will introduce uncorrelated vacuum noise that cannot be canceled by quantum destructive interference, leading to extra noise and reducing SNRs. This is not fundamental but poses practical challenges in device construction.

As we discussed at the end of Sects.\ref{sec:IVD} and \ref{sec:VIE}, high gain case in multi-stage SUI has not been treated theoretically because of the issue of non-commuting Hamiltonian of different PAs. High gain regime of parametric amplifier is important because it can generate EPR-type quantum entanglement in continuous variables \cite{reid,ou92}. It is the basis for complete quantum state teleportation and quantum metrology applications. Thus, this will be a challenge for future theoretical investigation of SUI.

SU(1,1) interferometers have the potential to reach the ultimate Heisenberg limit (HL) of phase measurement sensitivity. But as we have seen, internal losses are the main obstacle for improving sensitivity. How will losses affect the ability of SU(1,1) interferometers to reach Heisenberg limit?

The approach of SU(1,1) interferometer is to change the structures of interferometers by replacing beam splitters with parametric amplifiers. This is in contrast to the approach by using different quantum states for sensing. The former can be thought of as the hardware change whereas the latter as the software programming. The general condition for optimum quantum states was derived before \cite{ou97}. Is there an optimum interaction or hardware design in the construction of the non-traditional interferometers for sensing or other applications? Answer to this and other aforementioned questions will likely broaden our knowledge and applications of non-traditional interferometers.


\acknowledgments

The authors would like to thank the support by the National Natural Science Foundation of China (Grant Nos. 11527808, 91736105), the National Key Research and Development Program of China (Grant No. 2016YFA0301403), and the US National Science Foundation (Grant No. 1806425).
\vskip 0.2in

\noindent {\bf Data availability statement}

Data sharing is not applicable to this article as no new data were created or analyzed in this study.







\end{document}